\documentclass[english,tightenlines,10pt,eqsecnum,amsmath,amssymb,aps,prd,superscriptaddress,nofootinbib,showpacs,floats,notitlepage,showkeys]%
              {revtex4-1}
\usepackage{esint}
\usepackage{fmtcount}
\usepackage{tablefootnote}
\usepackage{threeparttable}
\usepackage{babel}
\usepackage{amstext,amsfonts}
\usepackage{epsfig}
\usepackage{color}
\usepackage{footnote}
\usepackage{amsmath}
\usepackage{fancyhdr}
\usepackage[dvipdfm]{hyperref}
\usepackage{hyperref}

\begin{document}
\title{Cosmological and Solar System Consequences of $f(R,T)$ Gravity Models}
\author{Hamid Shabani}\email{h\_shabani@sbu.ac.ir}
\author{Mehrdad Farhoudi}\email{m-farhoudi@sbu.ac.ir}
\affiliation{Department of Physics, Shahid Beheshti University,
             G.C., Evin, Tehran, 19839, Iran}
\date{August 7, 2014}
%
\begin{abstract}
\noindent To find more deliberate $f(R,T)$ cosmological solutions,
we proceed our previous paper further by studying some new aspects
of the considered models via investigation of some new
cosmological parameters/quantities to attain the most acceptable
cosmological results. Our investigations are performed by applying
the dynamical system approach. We obtain the cosmological
parameters/quantities in terms of some defined dimensionless
parameters that are used in constructing the dynamical equations
of motion. The investigated parameters/quantities are the
evolution of the Hubble parameter  and its inverse, the ``weight
function", the ratio of the matter density to the dark energy
density and its time variation, the deceleration, the jerk and the
snap parameters, and the equation--of--state parameter of the dark
energy. We numerically examine these quantities for two general
models $R+\alpha R^{-n}+\sqrt{-T}$ and $R\log{[\alpha
R]}^{q}+\sqrt{-T}$. All considered models have some inconsistent
quantities (with respect to the available observational data),
except the model with $n=-0.9$ which has more consistent
quantities than the other ones. By considering the ratio of the
matter density to the dark energy density, we find that the
coincidence problem does~not refer to a unique cosmological event,
rather, this coincidence also occurred in the early universe. We
also present the cosmological solutions for an interesting model
$R+c_{1}\sqrt{-T}$ in the non--flat FLRW metric. We show that this
model has an attractor solution for the late times, though with
$w^{(\textrm{DE})}=-1/2$. This model indicates that the spatial
curvature density parameter gets negligible values until the
present era, in which it acquires the values of the order
$10^{-4}$ or $10^{-3}$. As the second part of this work, we
consider the weak--field limit of $f(R,T)$ gravity models outside
a spherical mass immersed in the cosmological fluid. We have found
that the corresponding field equations depend on the both
background values of the Ricci scalar and the background
cosmological fluid density. As a result, we attain the
parametrized post--Newtonian (PPN) parameter for $f(R,T)$ gravity
and show that this theory can admit the experimentally acceptable
values of this parameter. As a sample, we present the PPN gamma
parameter for general minimal power law models, in particular, the
model $R+c_{1}\sqrt{-T}$.
\end{abstract}

\pacs{04.50.Kd; 95.36.+x; 98.80.-k; 98.80.Jk; 04.25.Nx}
 \keywords{Cosmology; $f(R,T)$ Gravity; Dark Energy;
 Dynamical Systems Approach; Modified Theories of Gravity;
 Solar System Solutions.}
\maketitle
\section{Introduction}\label{intro}
Up to the present observational data, the universe has undergone
an accelerated expansion phase~\cite{supno1,supno2,supno3}, the
explanation of which demands a theoretical paradigm based on an
ultimate theory. Until now, many authors have presented various
theories that are either new ones or those that are only
modified/generalized versions of the previous theories. For
example, one of the new theories is the string theory that
introduces some field theoretical version of general relativity
(GR) based on a new fundamental representation of matter called
``strings"~\cite{stri1,stri2}. Other theories of this type are
loop quantum gravity/cosmology~\cite{LQC1,LQG2,LQG3}, theories
based on the Ads/CFT correspondence~\cite{duality1,duality3} and
the holographic gravity/cosmology~\cite{holography1,holography2}.
In addition, there are higher--order gravities including the
special case $f(R)$ gravity~\cite{fR2,fR3,fR5,fR6,fR7,fR1,fR8},
the induced gravity~\cite{indu1,indu2}, the scalar--tensor
theories with the special case of Brans--Dicke
theory~\cite{brans1,brans3,brans4,brans5,brans6,brans2,brans7},
higher--dimensional theories, e.g., the Kaluza--Klein
theories~\cite{kaluza} and the braneworld scenarios~\cite{brane}.
Also, there are theories that introduce some modifications in the
matter component~\cite{visc1,visc2} or change the geometrical
structure, e.g., the non--commutative
theories~\cite{nonco1,nonco2,nonco3,nonco4,nonco5,nonco6}.

Most of these theories can be sorted in terms of two general
points of view related to the origin of the observed accelerated
expansion. On one hand, in some theories, this phenomenon is
explained by the impacts of a geometrical modification such as,
theories that take the dimension of space--time more than four,
theories that add some invariant geometrical scalars to the action
(e.g., $f(R)$ gravity), and theories that extract some idea from
quantum mechanics (e.g., the non--commutative cosmological
theories). And in some theories, the inhomogeneity of space--time
is responsible for this phenomenon~\cite{inhom1,inhom2}. On the
other hand, in some theories, some unknown fluid
 components are introduced to explain this problem. These
 unknown components are called
``dark energy''~\cite{fR3,dener1,dener2,dener3,dener4} which has a
contribution about $69\%$ of the  total matter
density~\cite{Planck} that accelerates the observed expansion of
the universe. There is also another exotic fluid called ``dark
matter"~\cite{dmatt1,dmatt2,dmatt3,dmatt4,dmatt5} that forms about
$26\%$ of the total matter density~\cite{Planck} and is
responsible for clustering of the galaxy structures. These two
puzzles are the main shortcomings of GR, for they are~not
predicted in this theory. These two contemporary observational
evidences have challenged our understandings about the universe.

There is a concordance model, namely the well--known $\Lambda$CDM
theory~\cite{LCDM}, which takes the Einstein--Hilbert action as
the geometrical sector and the dark energy and dark matter (in
addition to the usual baryonic matter) as the matter sectors. This
theory fits the present observational results, however, it has
some difficulties. In this theory, a cosmological constant plays
the role of dark energy, however, if this constant is pertained to
the vacuum energy, its problem will appear. This problem, which is
related to the non--comparability of the value of the dark energy
density with the field theoretical vacuum energy, is called the
``cosmological constant
problem"~\cite{Cos.pro1,Cos.pro2,Cos.pro3,Cos.pro4}. There has
been some impetus to tackle this problem in some theories, e.g.,
the dynamical dark energy models~\cite{Dyn.DE1,Dyn.DE2}.

One of the developments of $f(R)$ gravity is the idea introduced
in Ref.~\cite{frL1} that incorporates the matter Lagrangian
density together with an arbitrary function of the Ricci scalar as
an explicit non--minimal coupling. They have concluded that such a
combination leads to an extra force. This theory was progressed in
Refs.~\cite{frL2,frL3,frL4,frL5,frL6} and in Ref.~\cite{frL7}
which considers more complete forms. In recent years, a new theory
was developed in Ref.~\cite{harko}, named $f(R,T)$ gravity, that
can also be considered as a generalization of $f(R)$ gravity. In
this theory, an arbitrary function of the Ricci scalar and the
trace of the energy--momentum tensor is introduced instead of an
arbitrary function of only the Ricci scalar. The main
justifications for employing the trace of the energy--momentum
tensor may be the existence of some exotic matters or the
conformal anomaly (coming from the quantum
effects)\rlap.\footnote{See, e.g., Refs.~\cite{anomal1,anomal2}.}
The \textit{priori} appearance of the matter in an unusual
coupling with the curvature may also have some relations with the
issues such as geometrical curvature induction of matter, a
geometrical description of forces, and a geometrical origin for
the matter content of the universe\rlap.\footnote{See, e.g.,
Refs.~\cite{anomal2,fard} and references therein.} Since the
introduction of this theory, its numerous aspects have been
investigated, such as thermodynamics
properties~\cite{frttd1,frttd2,frttd3,frttd4}, energy
conditions~\cite{frtec1,frtec2,frtec3}, cosmological solutions
based on a homogeneous and isotropic space--time through a
phase--space analysis~\cite{phsp}, anisotropic
cosmology~\cite{frtani1,frtani2,frtani3}, a wormhole
solution~\cite{frtworm}, a cosmological solution via a
reconstruction program~\cite{frtrec1,frtrec2}, a cosmological
solution via an auxiliary scalar field~\cite{frtaf}, the study of
scalar perturbations~\cite{frtsp}, and some other
aspects~\cite{frtoth1,frtoth2,frtoth3}. Since for the
ultra--relativistic fluids, the trace of the energy--momentum
tensor vanishes, these components of matter do~not contribute in
the function of $f(R,T)$. To solve this lack, a generalization of
this theory has also been
established~\cite{frtnew1,frtnew2,frtec2}, in which a new
invariant, i.e., $R_{\alpha \beta}T^{\alpha \beta}$, has been
included.

In our previous paper~\cite{phsp}, we worked on the cosmological
solution of $f(R,T)$ gravity in a flat homogeneous and isotropic
background for a perfect fluid with zero equation--of--state
parameter via a phase space analysis. We investigated those
functions of $f(R,T)$ that can be decomposed in terms of a minimal
and/or a non--minimal combination of an arbitrary function of the
Ricci scalar, $g(R)$, and an arbitrary function of the trace of
the energy--momentum tensor, $h(T)$. Actually, the functions
$g(R)+h(T)$, $g(R)h(T)$ and $g(R)(1+h(T))$ were studied. We found
that the theories of the second type cannot have a consistent
cosmological solution. The investigation was based on the study of
some cosmological quantities, including the density parameters
(for the radiation, dust, and dark energy), the effective
equation--of--state parameter and the scale factor for the minimal
case of the theory. The corresponding diagrams of all these
quantities show, more or less, acceptable behaviors, hence one
cannot simply decide which one of these models is the best one.
Therefore, we should go through one more step and inspect these
models more accurately. Thus, in this work, we extend the previous
investigations further to consider more cosmological issues in the
minimal case. In this respect, in Sec.~\ref{Fieldequation}, we
briefly present the field equations of the theory and the related
definitions (as some of them were introduced in our previous
paper). Then, we report concisely the previous results in order to
reach a suitable connection to the issue. To check the consistency
of a theory with the observational data, all of the engaged
cosmological quantities should be considered. For this purpose, in
Sec.~\ref{newresults}, we probe some new cosmological quantities
including: the evolution of the Hubble parameter and its inverse
(which can be used as a loose estimation of the age or the size of
the universe), ``weight function" $g'(R)=d g(R)/d R$, the ratio of
the matter density to the dark energy density $r^{(\textrm{mD})}$
(hereafter, we call it the ``coincidence parameter") and its time
variation, the deceleration $q$, the jerk $j$ and the snap $s$
parameters, and the dark energy equation--of--state parameter
$w^{(\textrm{DE})}$. All these parameters/quantities are obtained
in terms of those dimensionless variables defined to reformulate
the equation of motions via the dynamical system procedure. One
will see that in $f(R,T)$ gravity, similar to $f(R)$ gravity, some
cosmological fluid densities are weighted by the function $F(R)$,
which implies that it is important to explore the behavior of this
function. This function must~not become negative, and, in the
matter--dominated era, it must be $F(R) \sim1$ in order to give GR
as a limiting solution. There is also a well--known dilemma called
``the coincidence problem". This problem deals with why the matter
and the dark energy densities are of the same order in the present
era. We do~not solve this problem, but we reach at the result that
this coincidence is~not a unique cosmological event and it had
also occurred in the early stages of the evolution of the
universe. Through the investigations, we present a number of
diagrams drawn numerically. In Sec.~\ref{simplecase}, we probe a
more plausible and simple model, namely,
$f(R,T)=R+c_{1}\sqrt{-T}$, in a non--flat background
space--time\rlap.\footnote{This model was~not considered in our
previous paper~\cite{phsp}.} This interesting model has some
similarity to the models with time--dependent cosmological
constant. One can interpret the term $\sqrt{-T}$ as a cosmological
constant depending on the matter that implicitly depends on time.
In the scaler--tensor models, like the dynamical dark energy and
the quintessence models~\cite{Dyn DE1,Dyn DE2}, in order to have
the late--time acceleration of the universe, extra components for
matter are introduced. However, the model $R+c_{1}\sqrt{-T}$
does~not contain any extra component for matter, and on the
contrary, it provides the late--time acceleration through the
interaction of the normal matter with the curvature. In
Sec.~\ref{weakfield}, we obtain the weak--field limit of $f(R,T)$
gravity outside a spherical body which is immersed in cosmological
fluid. We show that the cosmological fluid density plays an
important role in this theory. The PPN gamma parameter of this
theory is explicitly affected by the cosmological fluid density.
We show that this parameter can admit the experimentally accepted
values for $f(R,T)$ gravity. As an example, we obtain the PPN
parameter of the models $c_{\textrm{R}}R^{n}+c_{\textrm{T}}T^{m}$.
Finally, in Sec.~\ref{conclusions}, we summarize the results.
\section{Field equations and some primary consequences of
cosmological solutions of $f(R,T)=g(R)+h(T)$ gravity} \label{Fieldequation}
In this section, we briefly present the field equations of the
minimal case $f(R,T)=g(R)+h(T)$ gravity and also the dynamical
system structure of these equations, as introduced in
Ref.~\cite{phsp}. We do~not demonstrate the details, although we
mention the necessary information that makes us capable enough to
proceed further.

$f(R,T)$ gravity is introduced by the action
\begin{align}\label{action}
S=\int \sqrt{-g} d^{4} x \left[\frac{1}{16 \pi G} f(R,T^{\textrm{(m)}})+L^{\textrm{(total)}} \right],
\end{align}
where we have defined the Lagrangian of the total matter as
\begin{align}\label{total matter}
L^{\textrm{(total)}}\equiv L^{\textrm{(m)}}+L^{\textrm{(rad)}},
\end{align}
and $R$, $T^{\textrm{(m)}}\equiv g^{\mu \nu} T^{\textrm{(m)}}_{\mu
\nu}$, $L^{\textrm{(m)}}$ and $L^{\textrm{(rad)}}$ are the Ricci
scalar, the trace of the energy--momentum tensor of dust--like
matter, and the Lagrangians of the dust--like matter and
radiation, respectively. The superscript $m$ stands for the
dust--like matter, $g$ is the determinant of the metric and we set
$c=1$. The trace of the radiation energy--momentum tensor does~not
play any role in the function of $f(R, T^{\textrm{(m)}})$, for
$T^\textrm{(rad)}=0$, henceforth, we drop the superscript $m$ from
the trace $T^{\textrm{(m)}}$. The field equations are obtained as
\begin{align}\label{field tensor}
F(R,T) R_{\mu \nu}-\frac{1}{2} f(R,T) g_{\mu \nu}+\Big{(} g_{\mu \nu}
\square -\triangledown_{\mu} \triangledown_{\nu}\Big{)}F(R,T)=\Big{(}8
\pi G+ {\mathcal F}(R,T)\Big{)} T^{\textrm{(m)}}_{\mu \nu}+8 \pi G  T^\textrm{(rad)}_{\mu \nu},
\end{align}
where, for convenience, we have defined the following
functions for the derivatives with respect
to the trace $T$ and the Ricci scalar $R$. That is
\begin{align}
{\mathcal F}(R,T) \equiv \frac{\partial f(R,T)}{\partial T}=h'(T)~~~~~~~~~~\mbox{and}~~~~~~~~~~
F(R,T) \equiv \frac{\partial f(R,T)}{\partial R}=g'(R),
\end{align}
where the second equalities are for the considered minimal case
and the prime denotes the ordinary derivative with respect to
the argument. Contracting equation (\ref{field tensor}) gives
\begin{align}\label{trace tensor}
F(R,T)R+3 \square F(R,T)-2f(R,T)=\Big{(}8 \pi G+ {\mathcal F}(R,T)\Big{)}T,
\end{align}
where we have used
\begin{align}
g^{\alpha \beta} \frac{\delta T^{\textrm{(m)}}_{\alpha \beta}}{\delta g^{\mu \nu}}=-2T^{\textrm{(m)}}_{\mu \nu}
\end{align}
and the energy--momentum tensor is defined as
\begin{align}\label{Energy}
T_{\mu \nu}^{\textrm{(total)}}=-\frac{2}{\sqrt{-g}} \frac{\delta\left[\sqrt{-g} (L^{\textrm{(total)}})\right]}{\delta g^{\mu \nu}}.
\end{align}
Assuming that the total matter Lagrangian depends
only on the metric, the energy--momentum tensor reads
\begin{align}
T_{\mu \nu}^{\textrm{(total)}}=g_{\mu \nu} L^{\textrm{(total)}}-2\frac{\partial L^{\textrm{(total)}}}{\partial g^{\mu \nu}}.
\end{align}
We assume a perfect fluid and a spatially non--flat
Friedmann--Lema\^{\i}tre--Robertson--Walker (FLRW) metric
\begin{align}\label{metricFRW}
ds^{2}=-dt^{2}+a^{2}(t) \Big{(}\frac{dr^{2}}{1-kr^2}+r^{2}d\Omega^2\Big{)},
\end{align}
where $a(t)$ is the scale factor. To introduce the dark energy via
the effect of the extra term appearing from gravity modification,
we rewrite equation~(\ref{field tensor}) as the one that appears
in GR, i.e.,
\begin{align}\label{FRW}
G_{\mu \nu}=\frac{8 \pi G}{F(R,T)} \left(T^{\textrm{(total)}} _{\mu \nu}+T^{\textrm{(eff)}} _{\mu \nu}\right),
\end{align}
where
\begin{align}\label{Generalized einstein}
T^{\textrm{(eff)}} _{\mu \nu}\equiv \frac{1}{8 \pi G}\left[\frac{1}{2}
\Big{(}f(R,T)-F(R,T)R\Big{)}g_{\mu \nu}+\Big{(} \triangledown_{\mu}
\triangledown_{\nu}-g_{\mu \nu} \square\Big{)}F(R,T)+{\mathcal F}(R,T) T^{\textrm{(m)}}_{\mu \nu}\right].
\end{align}

In Ref.~\cite{phsp}, we have illustrated that the Bianchi identity
and the energy conservation law for the dust--like matter and
radiation independently lead to the constrain equation
\begin{align}\label{source}
\frac{3}{2} H(t) \mathcal F(R,T)=\dot {\mathcal F}(R,T),
\end{align}
which enforces us to choose a special form of function $f(R,T)$.
In equation (\ref{source}), the dot denotes the derivative with
respect to the cosmic time $t$. Equations (\ref{field tensor}) and
(\ref{trace tensor}), by assuming metric (\ref{metricFRW}), give
\begin{align}\label{first}
3H^{2}F(R,T)+\frac{1}{2} \Big{(}f(R,T)-F(R,T)R\Big{)}+3\dot{F}
(R,T)H+3\frac{kF(R,T)}{a^2}=\Big{(}8 \pi G +{\mathcal F}
(R,T)\Big{)}\rho^{\textrm{(m)}}+8 \pi G\rho^{\textrm{(rad)}}
\end{align}
as the Friedmann--like equation, and
\begin{align}\label{second}
2F(R,T) \dot{H}+\ddot{F} (R,T)-\dot{F} (R,T) H-2\frac{kF(R,T)}{a^2}=-\Big{(}8 \pi G
+{\mathcal F} (R,T)\Big{)}\rho^{\textrm{(m)}}-\frac{32}{3} \pi
G\rho^{\textrm{(rad)}}
\end{align}
as the Raychaudhuri--like equation.

The structure of phase space of the field equations in the minimal
case is simplified by defining a few variables and parameters.
These variables are generally defined as
\begin{align}
&x_{1}\equiv-\frac{\dot{F}}{H F},\label{varx1}\\
&x_{2}\equiv-\frac{g}{6 H^{2} F},\label{varx2}\\
&x_{3}\equiv \frac{R}{6 H^{2}}=\frac{\dot{H}}{H^{2}}+\frac{k}{H{^2}a^{2}}+2,\label{varx3}\\
&x_{4}\equiv-\frac{h}{3 H^{2} F},\label{varx4}\\
&x_{5}\equiv-\frac{T\mathcal{F}'}{3 H^{2} F},\label{varx5}\\
&\Omega^{\textrm{(rad)}}\equiv\frac{8\pi G \rho^{\textrm{(rad)}}}{3 H^{2} F},\label{omega rad}\\
&\Omega^{\textrm{(m)}}\equiv\frac{8\pi G \rho^{\textrm{(m)}}}{3 H^{2} F},\label{omega mat}\\
&\Omega^{\textrm{(k)}}\equiv-\frac{k}{H^{2}a^{2}},\label{omega cur}
\end{align}
and the parameters are
\begin{align}
&m \equiv \frac{R F'}{F},\label{parameterm}\\
&r \equiv -\frac{R F}{g}=\frac{x_{3}}{x_{2}}\label{parameterr},\\
&n \equiv \frac{T \mathcal{F}'}{\mathcal{F}}\label{n},\\
&s \equiv \frac{T\mathcal{F}}{h}=\frac{x_{5}}{x_{4}},\label{parameters}
\end{align}
where $R=6(\dot{H}+2H^2+k/a^2)$ for metric (\ref{metricFRW}). In
general, we have $m=m(r)$ and $n=n(s)$, and $g(R)\neq
\mbox{\textrm{constant}}$ and $h(T)\neq \mbox{\textrm{constant}}$.
Note that it is interesting that the spatial curvature density
parameter $\Omega^{\textrm{(k)}}$ does~not weight with the
function $F$ as the other density parameters do.

To relate the discussion of the dark energy to $f(R,T)$ modified gravity,
we redefine equations (\ref{first}) and (\ref{second}) as
\begin{align}\label{redefinition1}
3H^2F=8\pi G (\rho^{\textrm{(m)}}+\rho^{\textrm{(rad)}}+\rho^{\textrm{(k)}}+\rho^{\textrm{(DE)}})
\end{align}
and
\begin{align}\label{redefinition2}
-2F\dot{H}=8\pi G \Big{(}\rho^{\textrm{(m)}}+(4/3)\rho^{\textrm{(rad)}}+
(2/3)\rho^{\textrm{(k)}}+\rho^{\textrm{(DE)}}+p^{\textrm{(DE)}}\Big{)},
\end{align}
where $\rho^{\textrm{(k)}} \equiv -3kF/(8 \pi G a^{2})$, the density and the
pressure of the dark energy are defined as
\begin{align}\label{DEd}
8\pi G\rho^{\textrm{(DE)}}\equiv{\mathcal F}\rho^{\textrm{(m)}}-3\dot{F}(R,T)H-\frac{1}{2}\Big{(}f(R,T)-F(R,T)R\Big{)}
\end{align}
and
\begin{align}\label{DEp}
8\pi Gp^{\textrm{(DE)}}\equiv\ddot{F}(R,T)+2\dot{F}(R,T)H+\frac{1}{2}\Big{(}f(R,T)-F(R,T)R\Big{)}.
\end{align}
For the equation--of--state parameter of the dark energy, as
usual, we define $w^{\textrm{(DE)}}\equiv p^{\textrm{(DE)}} /
\rho^{\textrm{(DE)}}$.

By definitions (\ref{DEd}) and (\ref{DEp}), the continuity
equation for the dark energy is guarantied, i.e.,
\begin{align}\label{DEw}
\dot{\rho}^{\textrm{(DE)}}+3H\left(1+w^{\textrm{(DE)}}\right)\rho^{\textrm{(DE)}}=0.
\end{align}
Now, starting from the definition $w^{\textrm{(DE)}}$ and
replacing $\rho^{\textrm{(DE)}}$ and $p^{\textrm{(DE)}}$ from
(\ref{redefinition1}) and (\ref{redefinition2}) and then using the
definition of the effective equation of state as
$w^{\textrm{(eff)}}\equiv -1-2 \dot{H}/3 H^2$, the form
\begin{align}\label{weffd}
w^{\textrm{(eff)}}=\Omega^{\textrm{(DE)}} w^{\textrm{(DE)}}+
\frac{1}{3}\left(\Omega^{\textrm{(rad)}}-\Omega^{\textrm{(k)}}\right)
\end{align}
is obtained, where we have defined $\Omega^{\textrm{(DE)}}\equiv
8\pi G\rho^{\textrm{(DE)}}/3H^{2}F$. Finally, by definition
(\ref{varx3}), we obtain
\begin{align}\label{weff2}
w^{\textrm{(eff)}}=\frac{1}{3}\left(1-2x_{3}-2\Omega^{\textrm{(k)}}\right),
\end{align}
and by applying the conservation of the energy--momentum tensor and
equation (\ref{weff2}), for any perfect fluid, we get
\begin{align}\label{densityequation}
\dot{\rho}(t)+2\left(2-x_{3}-\Omega^{\textrm{(k)}}\right)H(t)\rho(t)=0.
\end{align}

The effect of constraint equation (\ref{source}) on the minimal
combination $f(R,T)=g(R)+h(T)$ restricts its form to a particular
one, namely,
\begin{align}\label{special form}
f(R,T)=g(R)+c_{1}\sqrt{-T}+c_{2},
\end{align}
where $c_{1}$ and $c_{2}$ are the integration constants. The
conservation of the energy--momentum tensor also leads to the case
in which the variable $x_{5}$ is a function of $x_{4}$, namely,
$x_{5}=x_{4}/2$, which in turn gives $s=-n=1/2$. These facts
simplify the later manipulations.

In the following, to avoid any complexity of the equations, we
discard the spatial curvature density, however, we will resume it
again in Sec.~\ref{simplecase}. Thus, we only present the
evolutionary equations of variables $x_{1}$ to $x_{4}$ and
$\Omega^{\textrm{(rad)}}$ and concisely discuss the cosmological
results. After some manipulations on the field equations
(\ref{first}) and (\ref{second}), we obtain the following
equations for the minimal case as
\begin{align}\label{eom1}
1+\frac{g}{6H^{2} g'} +\frac{h}{6 H^{2} g'}-\frac{R}{6 H^{2}} + \frac{\dot{g'}}
{H g'}=\frac{8 \pi G \rho^{\textrm{(m)}}}{3H^{2} g'} +\frac{h' \rho^{\textrm{(m)}}}
{3H^{2} g'}+\frac{8 \pi G \rho^{\textrm{(rad)}}}{3H^{2} g'}
\end{align}
and
\begin{align}\label{eom2}
2\frac{\dot{H}}{H^{2}}+\frac{\ddot{g'}}{H^{2} g'} -\frac{\dot{g'}}{H g'}=-\frac{8
\pi G \rho^{\textrm{(m)}}}{H^{2} g'}-\frac{h' \rho^{\textrm{(m)}}}{H^{2} g'}
-\frac{32\pi G \rho^{\textrm{(rad)}}}{3H^{2} g'}.
\end{align}
Equation (\ref{eom1}) gives the following constraint between
the matter and radiation density parameters and the variables $x_{1}$ to $x_{4}$, namely
\begin{align}\label{minimal matter density}
\Omega^{\textrm{(m)}} =1-\Omega^{\textrm{(rad)}}-x_{1}-x_{2}-x_{3}-x_{4}.
\end{align}
Also, by equation (\ref{minimal matter density}), we can define
the density parameter for the dark energy as
\begin{align}\label{Dark energy}
\Omega^{\textrm{(DE)}} \equiv x_{1}+x_{2}+x_{3}+x_{4},
\end{align}
hence, constraint (\ref{minimal matter density}) reads
\begin{align}\label{sum of densities}
\Omega^{\textrm{(DE)}}+\Omega^{\textrm{(m)}}+\Omega^{\textrm{(rad)}}=1.
\end{align}

Now, for the autonomous equations of motions, we obtain
\begin{align}
&\frac{d x_{1}}{d N}= -1+x_{1} (x_{1}-x_{3}) -3x_{2} -x_{3} -\frac{3}{2} {x_{4}}+\Omega^{\textrm{(rad)}} \label{minimal 1},\\
&\frac{d x_{2}}{d N}= \frac{x_{1} x_{3}}{m} +x_{2}\left(4+x_{1} -2x_{3}\right)\label{minimal 2},\\
&\frac{d x_{3}}{d N}=- \frac{x_{1} x_{3}}{m} +2x_{3} \left(2 -x_{3}\right)\label{minimal 3},\\
&\frac{d x_{4}}{d N}=x_{4}\left(\frac{5}{2}+x_{1}-2x_{3}\right)\label{minimal 4},\\
&\frac{d \Omega^{\textrm{(rad)}}}{d N}=\Omega^{\textrm{(rad)}}\left(x_{1}-2x_{3}\right)\label{minimal 5},
\end{align}
where $N=\ln a$ is used. This system of equations admits
ten fixed points including the following ones.
\begin{itemize}
\item Two points that have the characteristic of a
curvature--dominated point, and can lead to a stable
acceleration--dominated universe either in the non--phantom regime of
\begin{align}
&m'<-1,~~~~~~0<m<1,~~~~~~-1<w^{\textrm{(eff)}}<-\frac{1}{2},~~~\left(m\neq\frac{1}{2},
~~w^{\textrm{(eff)}}\neq-\frac{2}{3}\right),\\
&m'>-1,~~~~~~m<-\frac{1}{2}(1+\sqrt{3}),~~~~~~-1<w^{\textrm{(eff)}}<-\frac{1}{3}
\end{align}
or in the phantom regime of
\begin{align}
&m'>-1,~~~~~~m>1,~~~~~~-1.07<w^{\textrm{(eff)}}<-1,\\
&m'>-1,~~~~~~-\frac{1}{2}<m<0,~~~~~~w^{\textrm{(eff)}}<-7.60.
\end{align}
\item One saddle matter--dominated era point that exists, provided that $m\rightarrow 0^{+}$.
\item One stable de Sitter point that exists for $0<m<1$.
\end{itemize}

Based on the existence of these fixed points, there are six
classes of the cosmological solutions that pass a long--enough
matter--dominated era followed by an accelerated expansion. In
four classes, the universe approaches a state with a phantom or
non--phantom dark energy, and in the other two classes, the
universe approaches a de Sitter point. The rest of the fixed
points are~not physically interested. For more details of this
classification, one can refer to Ref.~\cite{phsp}. Also, the main
features of the solutions are as follows:
\begin{itemize}

\item An interesting feature in these solutions is
the appearance of a solution with a transient period
of acceleration with $w^{\textrm{(eff)}}=-1/2$
followed by a de Sitter acceleration as the final
attractor, see the diagram for $w^{\textrm{(eff)}}$ in
Figure~\ref{CaseC1}. With different initial values,
one can obtain diagrams with a short--enough transient
era to reach a state with $w^{\textrm{(eff)}}=-1$ in order to
match the present observations. For example,
this feature appears in the model with
$g(R)=R+\alpha R^{0.9}$, see Figure~\ref{CaseE2}.
\end{itemize}
\begin{itemize}
\item All the models considered in Ref.~\cite{phsp} show
a satisfying sequence of radiation--matter--acceleration
era, similar to the one in Figure~\ref{CaseE2}.
\end{itemize}

The above features, more or less, appeared in all the models
already considered in Ref.~\cite{phsp}, which indicates that one
should investigate further some other cosmological aspects of
these models to determine the more compatible ones to the
observational data. In this respect, in the following section, we
numerically discuss the coincidence problem and some other
cosmological parameters for only two more interesting models of
Ref.~\cite{phsp} case by case since (based on our checks) the
other models show the same properties.
\section{New aspects of cosmological solutions of $f(R,T)=g(R)+h(T)$ gravity}\label{newresults}
As emphasized in the Introduction, an acceptable theory must
contain the well--behaved cosmological parameters/quantities that
match the available observational data. Otherwise, theories with
inconsistent parameters/quantities will be mostly ruled out. In
this section, we discuss some new cosmological
parameters/quantities for two of the theories introduced in our
previous work~\cite{phsp}, the other theories considered in
Ref.~\cite{phsp} have similar properties, and thus to avoid
repetitious results and mathematics, we do~not investigate these
theories.

The values of energy densities of the matter and dark energy
components are, up to the cosmological observations, of the same
order in the present epoch. This evidence has given rise to a
famous dilemma called the coincidence problem. This problem deals
with a question, that is, why the ratio of the matter density to
the dark energy density is of the same order at the present era.
In addition to the coincidence problem, we present some results on
the Hubble parameter, the weight function $F$, the deceleration
parameter, the jerk and the snap. The importance of the function
$F$ first is that it does~not attain negative values, for the
definitions of the density parameters (\ref{omega rad}) and
(\ref{omega mat}) are weighted by this function. Actually,
negative values of this function can comprise the consequence of
repulsive gravity. Second, we must have $F(R)\sim 1$ in order to
get GR as a limiting solution.

Regarding the coincidence problem, we define a coincidence
parameter as the ratio of the matter energy density to the dark
energy density, namely,
\begin{align}\label{rho-mD1}
r^{\textrm{(mD)}}\equiv \frac{\rho^{\textrm{(m)}}}{\rho^{\textrm{(DE)}}}=
\frac{\Omega^{\textrm{(m)}}}{\Omega^{\textrm{(DE)}}}.
\end{align}
Using equations (\ref{minimal matter density})
and (\ref{Dark energy}), definition
(\ref{rho-mD1}) can be rewritten as
\begin{align}\label{rho-mD2}
r^{\textrm{(mD)}}=\frac{1-\Omega^{\textrm{(rad)}}-x_{1}-x_{2}-x_{3}-x_{4}}{x_{1}+
x_{2}+x_{3}+x_{4}}.
\end{align}
Differentiating it with respect to $N$ gives
\begin{align}\label{rho-mD2 dot1}
\frac{d r^{\textrm{(mD)}}}{dN}=3r^{\textrm{(mD)}}w^{\textrm{(DE)}}.
\end{align}
Also, from equation (\ref{weffd}), one can obtain
\begin{align}\label{weffy}
w^{\textrm{(DE)}}=\frac{1-2x_{3}-\Omega^{\textrm{(rad)}}-\Omega^{\textrm{(k)}}}{3\left(x_{1}+x_{2}+x_{3}
+x_{4}\right)}.
\end{align}
Hence, one can obviously
have relation (\ref{rho-mD2 dot1}) in terms of the dimensional
variables.

On the other hand, for the Hubble parameter, we can also obtain an
equation in terms of the dimensionless variables. Let us rewrite
definition (\ref{varx3}) as
\begin{align}\label{varx3-re}
H=\sqrt{\frac{R}{6x_{3}}}
\end{align}
and from definition (\ref{parameterr}) define $R$  as a function
of $r$, i.e.,
\begin{align}\label{R-function}
R=Q(r)=Q(\frac{x_{3}}{x_{2}}),
\end{align}
where, in general, $Q(r)$ is obtained by solving equation
(\ref{parameterr}) for a given function $g(R)$. Thus, by relations
(\ref{varx3-re}) and (\ref{R-function}), the Hubble parameter can,
in principle, be rewritten in terms of the dimensionless
variables, namely,
\begin{align}\label{H-function}
H(x_{2},x_{3})=\sqrt{\frac{Q(\frac{x_{3}}{x_{2}})}{6x_{3}}}.
\end{align}
Also, the function $F=g'(R)$ can be cast in the form
of $F(x_{2},x_{3})$ by employing
relation (\ref{R-function}).

Now, the other cosmological parameters, namely, the deceleration
parameter, the jerk, and the snap, are usually defined via the
expansion of an analytical scale factor around its present value
$a(t_{0})\equiv a_{0}$ , i.e.,
\begin{align}\label{Scale-expand}
\frac{a(t)}{a_{0}}=1+H_{0}(t-t_{0})-\frac{1}{2}q_{0}H_{0}^{2}(t-t_{0})^{2}+
\frac{1}{6}j_{0}H_{0}^{3}(t-t_{0})^{3}+\frac{1}{24}s_{0}H_{0}^{4}(t-t_{0})^{4}+\mathcal{O}[(t-t_{0})^{5}],
\end{align}
where these dimensionless parameters are defined as
\begin{align}\label{deceleration}
q(t)\equiv-\frac{\ddot{a}}{a}H^{-2}
\end{align}
for the deceleration parameter,
\begin{align}\label{jerk}
j(t)\equiv\frac{\dddot{a}}{a}H^{-3}
\end{align}
for the jerk parameter, and
\begin{align}\label{snap}
s(t)\equiv\frac{\ddddot{a}}{a}H^{-4}\end{align} for the snap
parameter. From definition (\ref{deceleration}) and the definition
for the effective equation of state $w^{\textrm{(eff)}}$, the
deceleration parameter is
\begin{align}\label{q-weff}
q=-(1+\frac{\dot{H}}{H^{2}})=\frac{1}{2}\left[1+3w^{(\textrm{eff})}\right].
\end{align}
However, one can also rewrite it in terms of the defined
dimensionless variable as
\begin{align}\label{q-x3}
q=1-x_{3}.
\end{align}
Using equations (\ref{deceleration}), (\ref{jerk}), and
(\ref{q-weff}), the jerk is
\begin{align}\label{j-q}
j=q(1+2q)-\frac{dq}{dN}
\end{align}
or, in terms of the dimensionless variables (\ref{varx1}),
(\ref{varx3}), and (\ref{parameterm}), is
\begin{align}\label{j-xi}
j=-\frac{x_{1}x_{3}}{m}-x_{3}+3.
\end{align}
Similar calculations can be performed for the snap parameter,
namely,
\begin{align}\label{s-jq}
s=-j(2+3q)+\frac{dj}{dN},
\end{align}
and then
\begin{align}\label{s-xi}
s=\frac{x_{1}x_{3}}{m^{2}}\left(x_{1}+\frac{dm}{dN}\right)+
\frac{1}{m}\left[x_{3}\left(1-x_{1}(x_{1}-2)+3x_{2}+x_{3}+\frac{3}{2} x_{4}-\Omega^{(\textrm{rad})}\right)\right]
-x^{2}_{3}+10x_{3}-15.
\end{align}
It is interesting to note that, except the deceleration parameter,
both the jerk and the snap parameters explicitly depend on the
parameter $m$ and therefore to the chosen model. By using the
value of $w^{\textrm{(eff)}}$ for different cosmological eras and
the above equations, the values of the corresponding parameters
are shown in Table~\ref{Tab1}.

By choosing suitable initial values
and parameters, we will investigate the above results for the
following two theories in the subsequent subsections.

\begin{center}
\begin{table}[h]
\centering
\caption{The values of the deceleration, jerk
and snap cosmological parameters.}
\begin{tabular}{l @{\hskip 0.4in} c@{\hskip 0.4in} c @{\hskip 0.4in}c}\hline\hline

Cosmological eras     &Deceleration parameter q &Jerk parameter j  &Snap parameter s\\[0.5 ex]
\hline
Radiation domination&$1$&$3$&$-15$\\[0.75 ex]
Matter domination&$\frac{1}{2}$&$1$&$-\frac{7}{2}$\\[0.75 ex]
De Sitter domination&$-1$&$1$&$1$\\[0.75 ex]
\hline\hline
\end{tabular}
\label{Tab1}
\end{table}
\end{center}
\subsection{$f(R,T)= R^{p}[\log{(\alpha R})]^q+\sqrt{-T}$,~~~$q\neq0$,~~$\alpha>0$}\label{sub1}
This Lagrangian, for $p=1$ and $q=1$, leads to a model with
\begin{align}\label{F(x)-A}
F(x_{2},x_{3})=\frac{x_{3}}{x_{2}+x_{3}}~~~~~~~~~~\mbox{and}~~~~~~~~~~
H(x_{2},x_{3})=\sqrt{\frac{\exp{[-(\frac{x_{2}}{x_{2}+x_{3}})}]}{6\alpha
x_{3}}}
\end{align}
and, for $p=1$ and $q=-1$, to
\begin{align}\label{F(x)-A}
F(x_{2},x_{3})=-\frac{x_{3}}{x_{2}^2}(x_{2}+x_{3})~~~~~~~~~~\mbox{and}~~~~~~~~~~
H(x_{2},x_{3})=\sqrt{\frac{\exp{(\frac{x_{2}}{x_{2}+x_{3}})}}{6\alpha x_{3}}}.
\end{align}

\begin{figure}[h]
\epsfig{figure=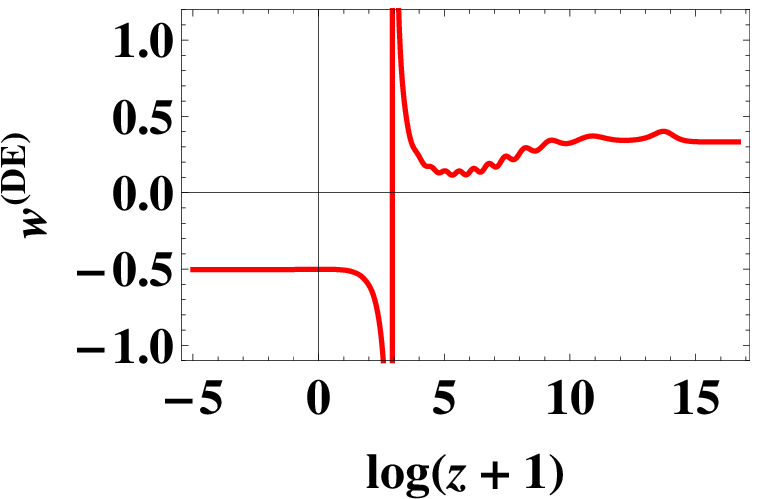,width=4.25cm}\hspace{2mm}
\epsfig{figure=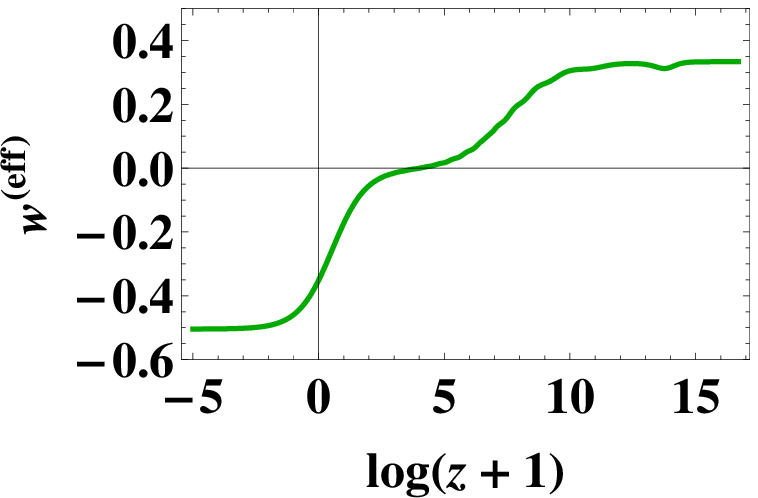,width=4.25cm}\hspace{2mm}
\epsfig{figure=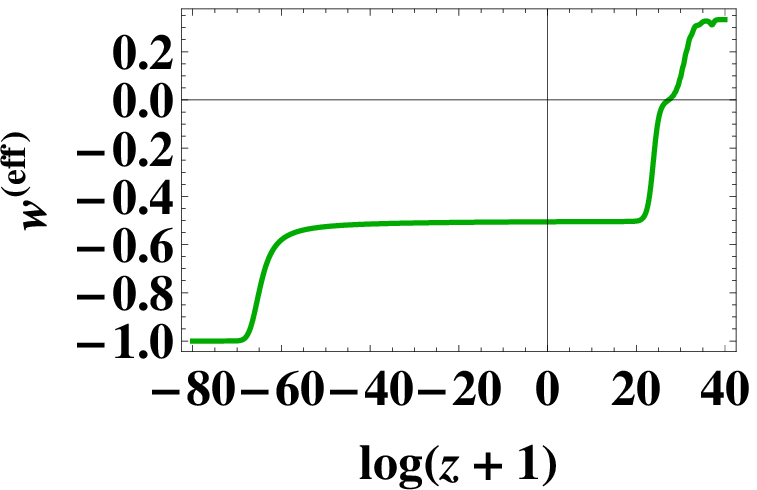,width=4.25cm}\hspace{2mm}
\epsfig{figure=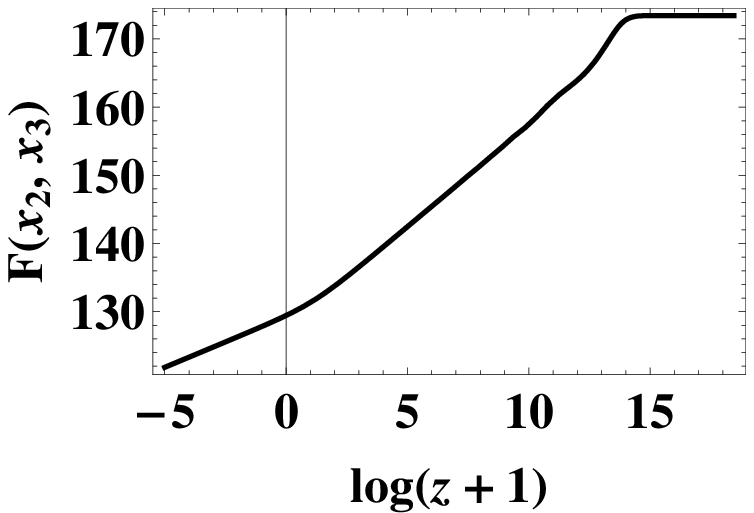,width=4.25cm}\vspace{2mm}
\epsfig{figure=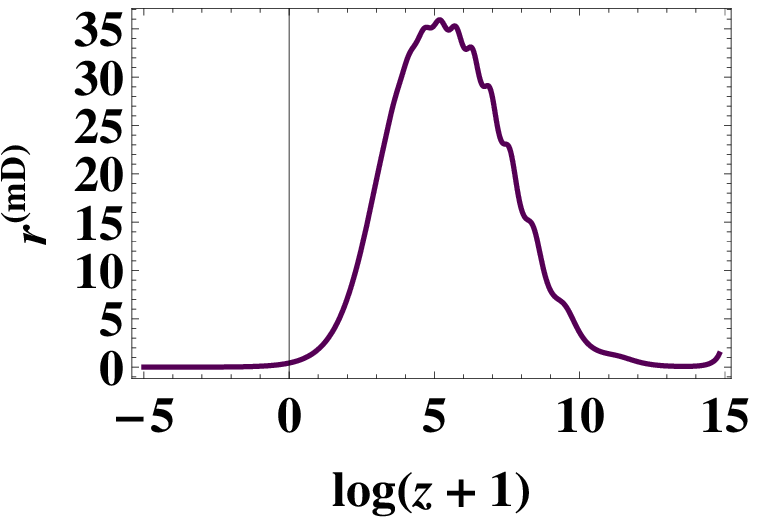,width=4.25cm}\hspace{2mm}
\epsfig{figure=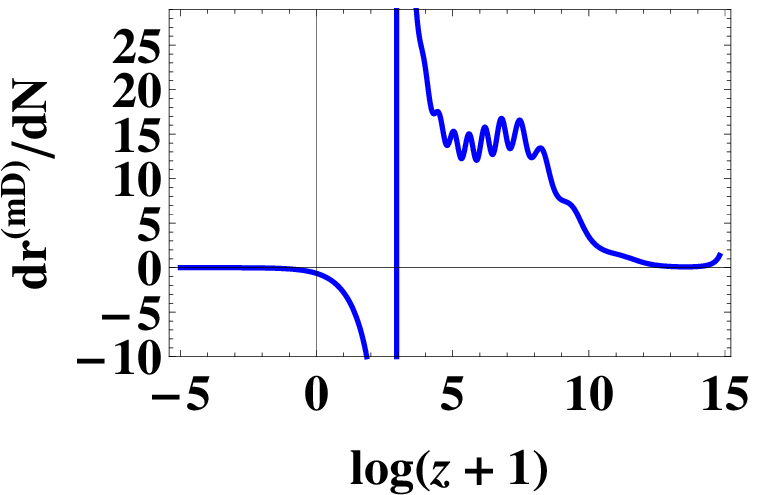,width=4.25cm}\hspace{2mm}
\epsfig{figure=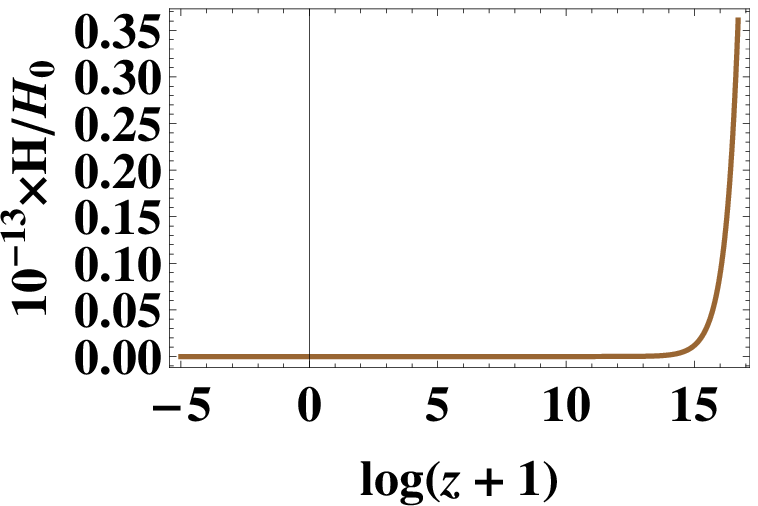,width=4.25cm}\hspace{2mm}
\epsfig{figure=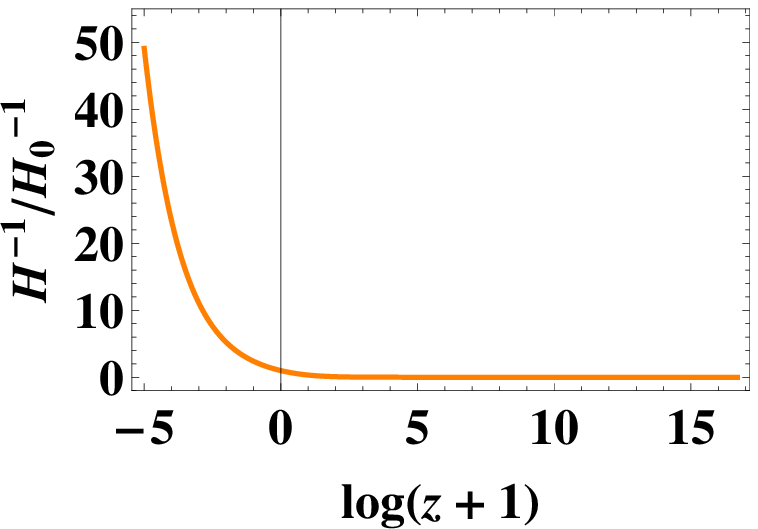,width=4.25cm}\ \caption{(color
online).\footnotesize {\textbf{ Cosmological solutions of
$f(R,T)=R\log{(\alpha R)}+\sqrt{-T}$ gravity.}} The different
cosmological parameters consist of the dark energy equation of
state, the effective equation of state, $F(x_{2},x_{3})$, the
ratio of the matter density to the dark energy density and its
first derivative, and the Hubble parameter and its inverse. The
initial values $x_{1}=10^{-7}$, $x_{2}=-10^{-6}$, $x_{3}=1.0058
\times 10^{-6}$, $x_{4}=0.3 \times 10^{-14}$, and $x_{5}=0.9999$
correspond to $z\thickapprox 1.78 \times 10^{7}$. We used $\alpha
= 1.95 \times 10^{92}$ in some plots. The diagrams are drawn to be
consistent with the present value of $\Omega^{\textrm{(m)}}
_{\textrm{0}}\approx0.3$, $\Omega^{\textrm{(rad)}}_{\textrm{0}}
\approx10^{-4}$ and $H_{0}\simeq 67.3 km/(Mpc.s)$. As it is
obvious, the diagram for $F(x_{2},x_{3})$ has~not a suitable range
of value to indicate a matter--dominated era for which $F \approx
1$. The diagram of $r^{\textrm{(mD)}}$ shows that the value of
$3/7$ appears twice in the evolution of the cosmos, and its ratio
is about $-0.6$ at the present era. The matter dominates in the
redshift $z\sim 33$, and from the diagram for $H^{-1}$, we find
that in this epoch the relative size of the universe is about $9.4
\times 10^{-3}$.} \label{CaseC1}
\end{figure}

\begin{figure}[h]
\epsfig{figure=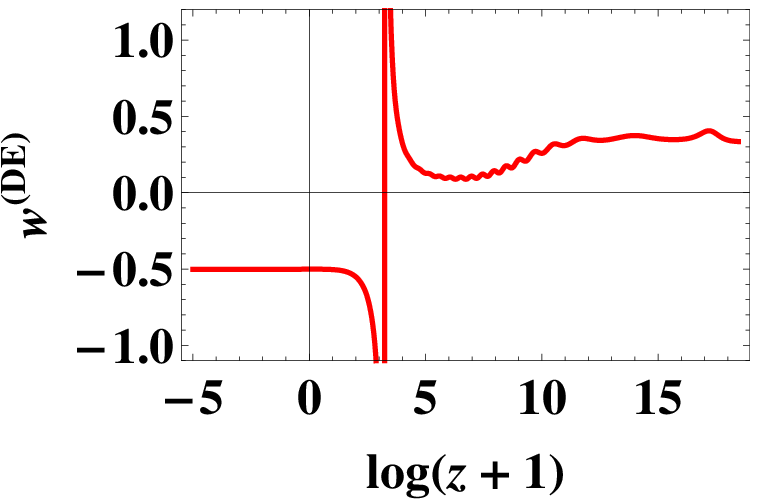,width=4.25cm}\hspace{2mm}
\epsfig{figure=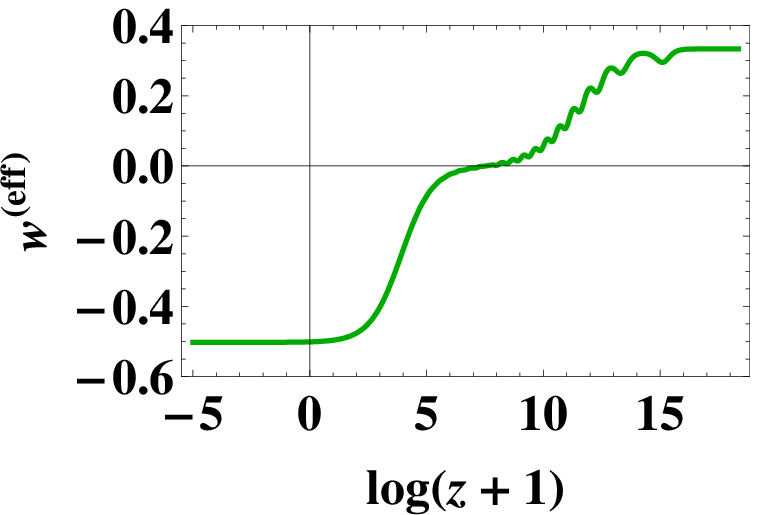,width=4.25cm}\hspace{2mm}
\epsfig{figure=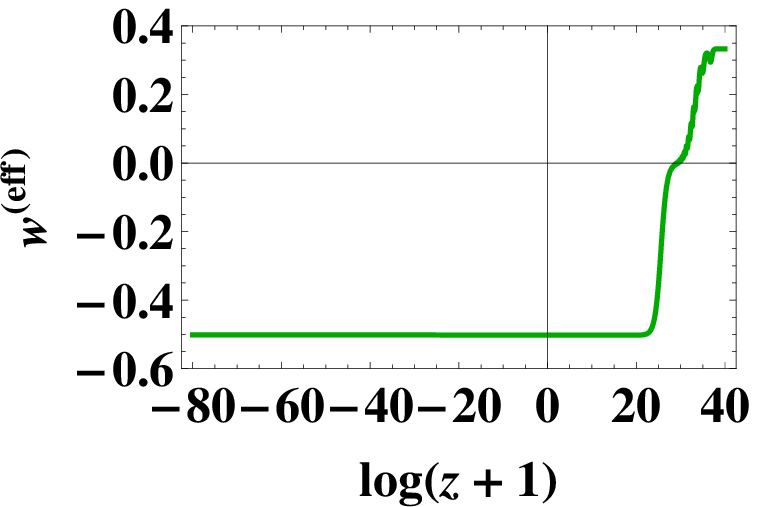,width=4.25cm}\hspace{2mm}
\epsfig{figure=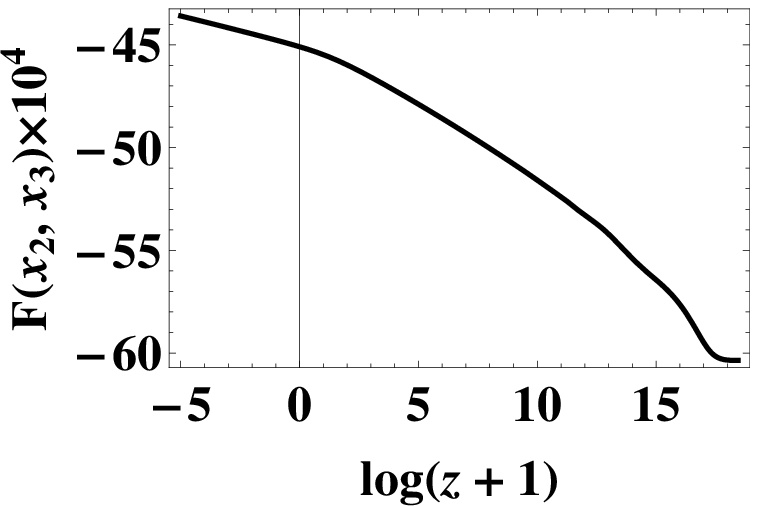,width=4.25cm}\vspace{2mm}
\epsfig{figure=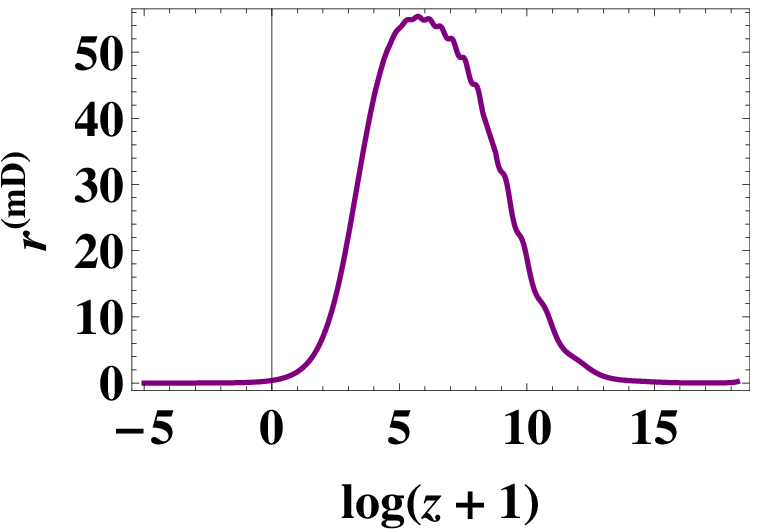,width=4.25cm}\hspace{2mm}
\epsfig{figure=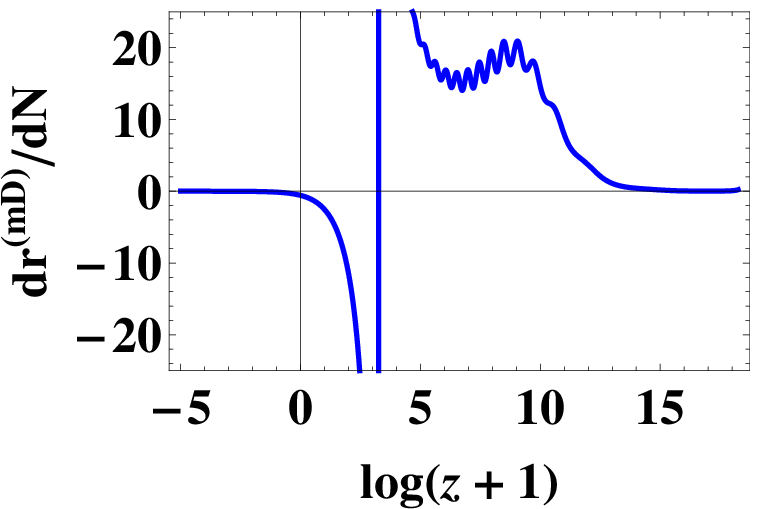,width=4.25cm}\hspace{2mm}
\epsfig{figure=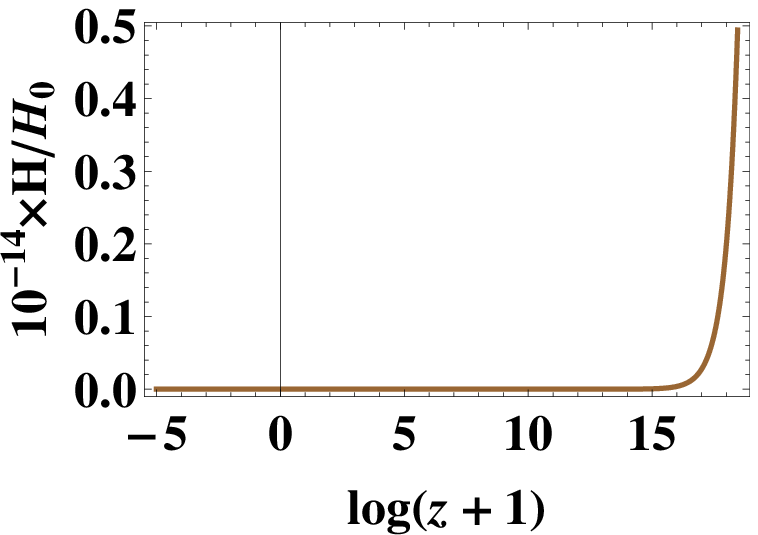,width=4.25cm}\hspace{2mm}
\epsfig{figure=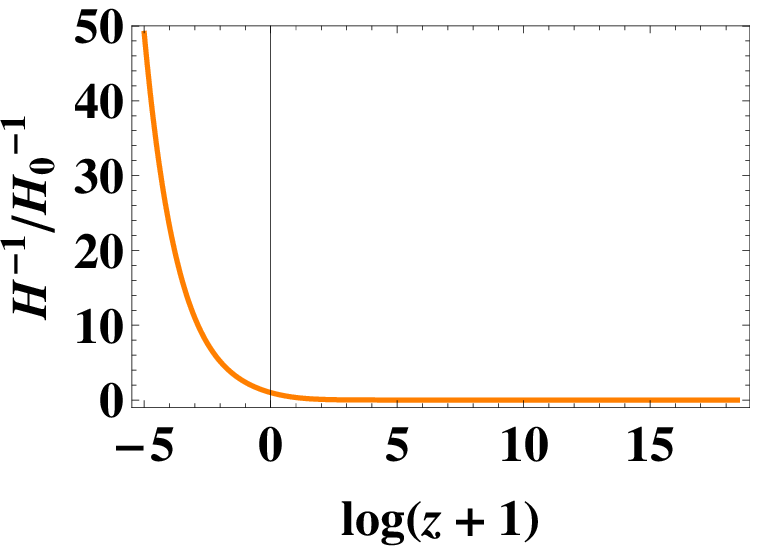,width=4.25cm}\ \caption{(color
online).\footnotesize {\textbf{ Cosmological solutions of
$f(R,T)=R/ \log{(\alpha R)}+\sqrt{-T}$ gravity.}} The initial
values are $x_{1}=10^{-9}$, $x_{2}=-10^{-3}$, $x_{3}=1.006 \times
10^{-3}$, $x_{4}=2.3 \times 10^{-16}$ and $x_{5}=0.9999$
corresponding to $z\thickapprox 1.03 \times 10^{8}$. We used
$\alpha = 5.7 \times 10^{61}$ in some plots. The numerical data
are matched to the present values of
$\Omega^{\textrm{(m)}}_{\textrm{0}}\approx0.3$,
$\Omega^{\textrm{(rad)}}_{\textrm{0}}\approx10^{-4}$ and
$H_{0}\simeq 67.3 km/(Mpc.s)$. In this case, the values of
$F(x_{2} ,x_{3})$ are negative, which is~not accepted. The
matter--dominated era happens in the redshift $z\sim 60$, and the
relative size of the universe in this era is about $4.1 \times
10^{-3}$.} \label{CaseC2}
\end{figure}

In Figure~\ref{CaseC1}, we present the related diagrams for the
case $p=1$ and $q=1$. When $\alpha=1.95\times10^{92}$, the
diagrams for the Hubble parameter and its inverse are also
presented. For $q=-1$ with  $\alpha=5.7\times10 ^{-61}$, we have
drawn the diagrams in Figure~\ref{CaseC2}. Note that, in both
cases, $\alpha$ has extraordinary large and small values (in the
dimension of length squared), which reminds us of some kind of
fine--tuning problem. In the model with $q=1$, the universe passes
a transient period of accelerated expansion with
$w^{(\textrm{eff})} =-1/2$ and then experiences a stable de Sitter
era with $w^{(\textrm{eff})} =-1$, though the model with $q=-1$
accepts an accelerated expansion with $w^{(\textrm{eff})} =-1/2$
in the late times. The diagram of dark energy for both models
exhibits a singular point that is a known behavior in $f(R)$
gravities. A caution is needed for the function $F(R)$ that
appears in the denominator of $\Omega^{(\textrm{m})}$ and
$\Omega^{(\textrm{rad})}$. First, this function must be a positive
value, and second, it should be of the order of $1$ to be
consistent with the matter era solution for which $g(R)\sim R$. It
means that one should recover the matter era solution for which
$F(R)=1$. The model with $q=1$ gets positive values in all times,
though, the attained values are far from $1$. The other model
gives negative values, which is~not an interesting result. For the
coincidence parameter $r^{\textrm{(mD)}}$ and its first
derivative, the diagrams are drawn in Figure~\ref{CaseC1}. An
interesting feature is that this ratio has a peak, which means the
dark energy density parameter increases after the domination
period of matter, and this ratio gets the value
$r^{\textrm{(mD)}}\simeq 3/7$, twice in the history of cosmology.
The first one occurs in the outset of deceleration period (which
means before this time, the dark energy density parameter has been
dominated over the matter density parameter), when the matter
density exceeds the dark energy density. The other one happens in
the outset of acceleration when the dark energy density exceeds
the matter density. In this case, we have $dr^{\textrm{(mD)}}/dN
\simeq -0.6$ in the present era, which shows $r^{\textrm{(mD)}}$
increases up to zero. The Hubble diagram and the Hubble radius are
also presented in Figure~\ref{CaseC1}. The resulted universe is
small until the acceleration era. In the theory with $q=1$, the
matter domination takes place in the redshift $z\sim 33$ and the
relative size of the universe\footnote{By the relative size of the
universe, we mean the ratio $cH^{-1}(z_{m})/cH^{-1}(z_{0})$ in a
loose estimation, where $z_{m}$ is the redshift in which the
matter dominates and $z_{0}$ determines its present value.} in
this epoch, with respect to its present value, is about $9.4
\times 10^{-3}$, which are fully inconsistent with the available
data~\cite{mukhanov}. For the theory with $q=-1$, the
corresponding values are $z\sim 60$ and $4.1 \times 10^{-3}$. Note
that different initial values may improve these values.
\begin{center}
\begin{table}[t]
\centering
\caption{Different features of $f(R,T)$ cosmological models.}
\begin{minipage}{16cm}
\begin{tabular}{l @{\hskip 0.15in} l@{\hskip 0.15in} l @{\hskip 0.15in}l
@{\hskip 0.15in} l @{\hskip 0.15in} l @{\hskip 0.2in} l @{\hskip 0.15in}
l @{\hskip 0.15in} l @{\hskip 0.15in} l}\hline\hline

$f(R,T)$ Model     &$w^{\textrm{(DE)}}_{l}$\footnote{``$l$'' denotes the late--time values.}    &$H^{-1}_{0}$\footnote{``$0$'' denotes the present values.}    &$F^{\textrm{(m)}}\footnote{The superscript $m$ indicates the value of parameter in the matter--dominated era.}$    &$(\frac{dr}{dN}^\textrm{(mD)})_{0}$    &$\alpha$    &$z^{\textrm{(m)}}$   &$q_{0}$     &  $j_{0}$      &$s_{0}$\\[0.5 ex]
\hline
$R\log{(\alpha R)}+\sqrt{-T}$    &$-1$            &$9.3 \times 10^{-3}$   &138     &-0.64   &$1.95\times10^{92}$   &32.6   &-0.027   &0.21   &-0.76   \\[0.75 ex]
$R/\log{(\alpha R)}+\sqrt{-T}$   &$-\frac{1}{2}$  &$4.1 \times 10^{-3}$   &-0.005  &-0.58   &$5.7\times10^{-61}$   &58.7   &-0.251   &-0.12   &0.18   \\[0.75 ex]
$R+\alpha R^{1.01}+\sqrt{-T}$  &$-\frac{1}{2}$  &$5.9 \times 10^{-3}$   &1.66    &-0.43   &$1.371$               &43.5   &-0.016   &0.23   &-0.80   \\[0.75 ex]
$R+\alpha R^{0.9}+\sqrt{-T}$   &$-1$            &$1.2 \times 10^{-5}$   &0.97    &-0.68   &$0.646\times10^{-4}$  &1104   &-0.038   &0.22   &-0.76   \\[0.75 ex]
\hline\hline
\end{tabular}
\end{minipage}
\label{Tab2}
\end{table}
\end{center}

\subsection{$f(R,T)= R+\alpha R^{-n}+\sqrt{-T}$,~~~$n\neq0$}\label{modelC}
\begin{figure}[h]
\epsfig{figure=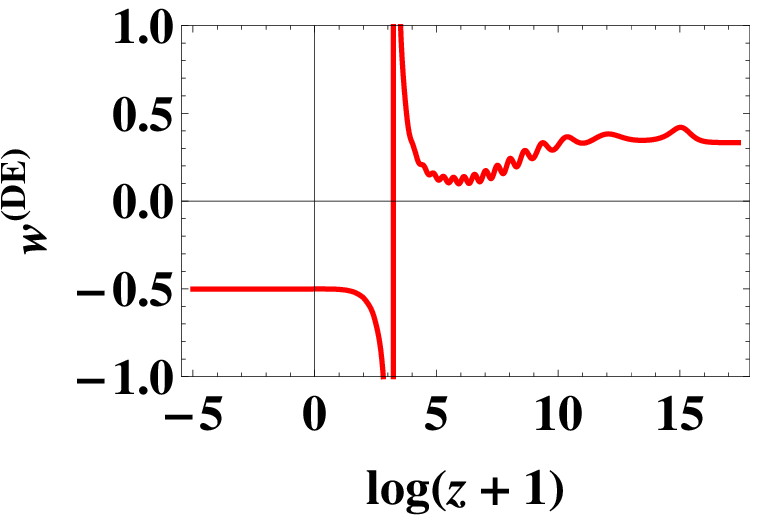,width=4.25cm}\hspace{2mm}
\epsfig{figure=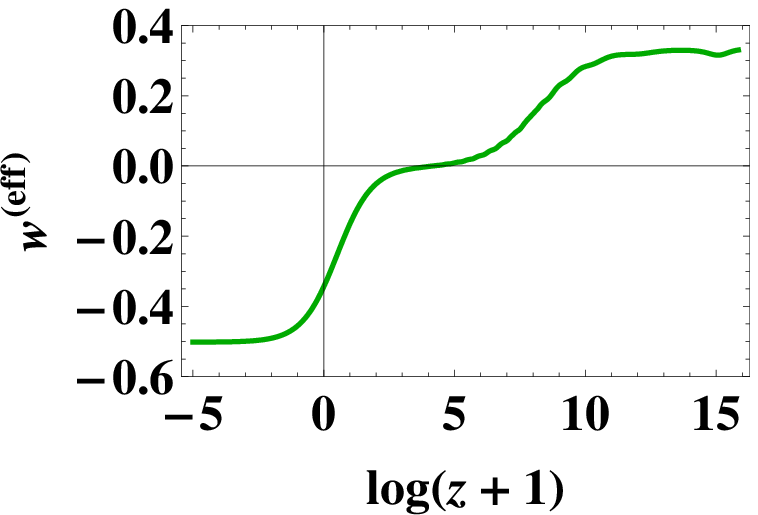,width=4.25cm}\hspace{2mm}
\epsfig{figure=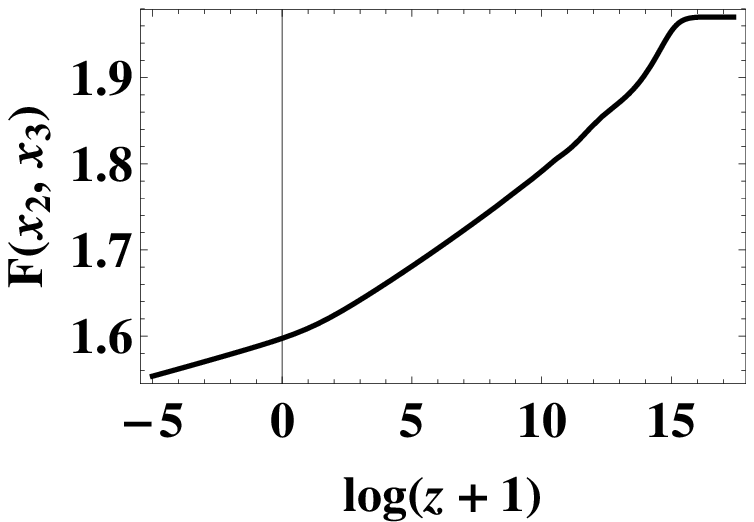,width=4.25cm}\hspace{2mm}
\epsfig{figure=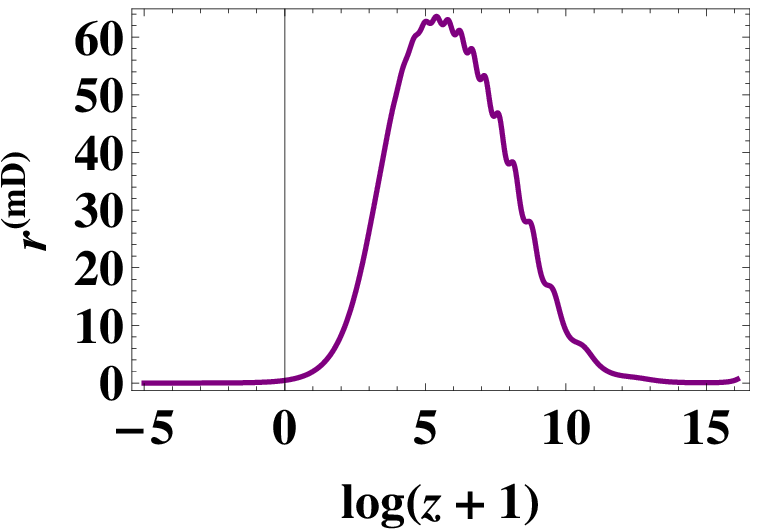,width=4.25cm}\vspace{2mm}
\epsfig{figure=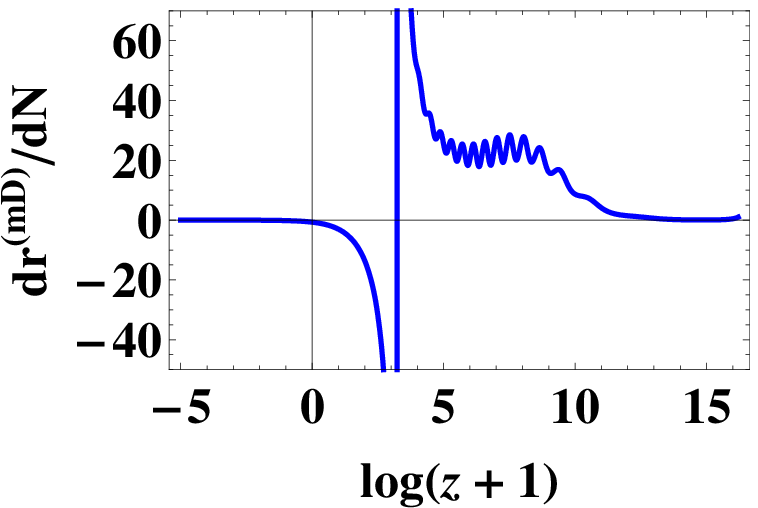,width=4.25cm}\hspace{2mm}
\epsfig{figure=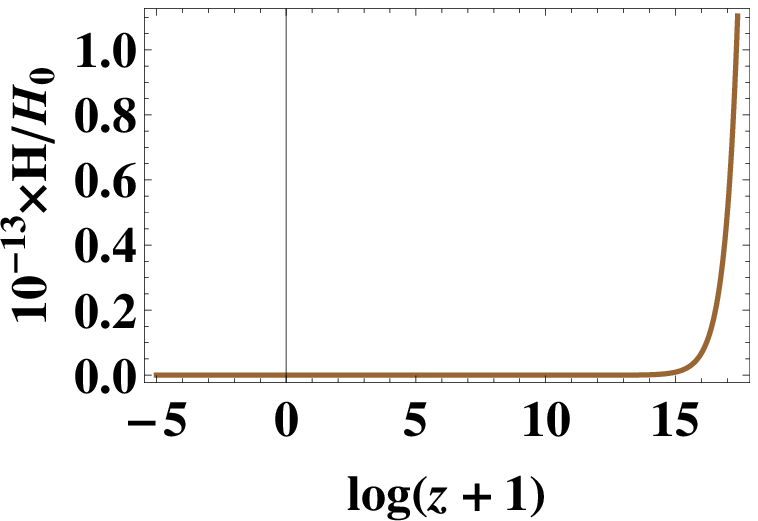,width=4.25cm}\hspace{2mm}
\epsfig{figure=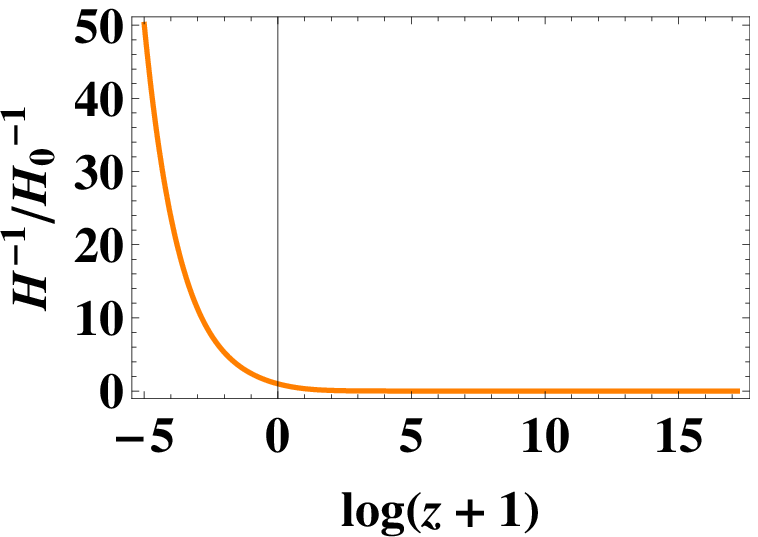,width=4.25cm}\ \caption{(color
online).\footnotesize {\textbf{ Cosmological solutions of
$f(R,T)=R+ \alpha R^{1.01}+\sqrt{-T}$ gravity.}} In this case, the
initial values $x_{1}=10^{-8}$, $x_{2}=-10^{-5}$, $x_{3}=1.0049
\times 10^{-5}$, $x_{4}=10^{-15}$, and $x_{5}=0.9999$ correspond
to $z\thickapprox 3.6 \times 10^{7}$. We have used $\alpha = 1.43$
in some plots. In this theory, we reach at the same results as the
previous two theories in subsection \ref{sub1}, except the result
of the function $F$, which is more reasonable in this theory. The
matter--dominated era for the theory with $n=-1.01$ occurs in
$z\sim 43$ with the relative size of the universe in this era
about $6 \times 10^{-3}$.} \label{CaseE1}
\end{figure}

\begin{figure}[h]
\epsfig{figure=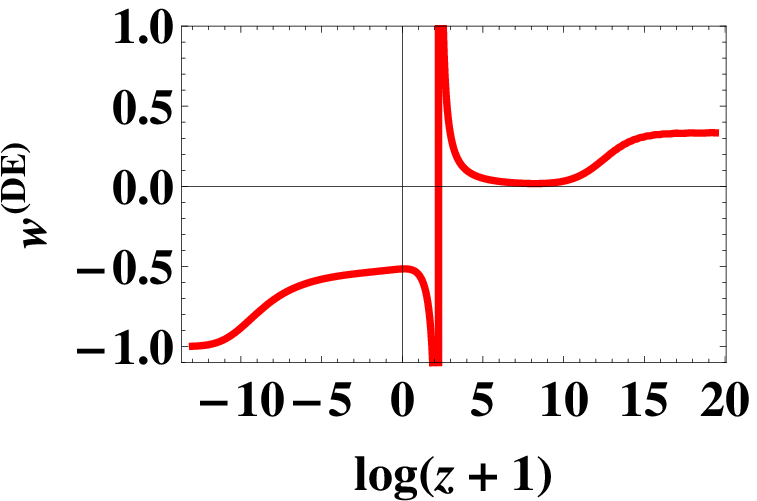,width=4.25cm}\hspace{2mm}
\epsfig{figure=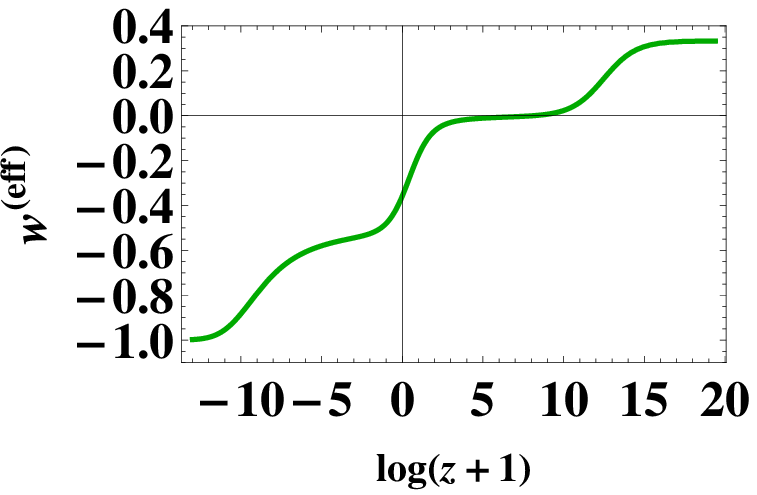,width=4.25cm}\hspace{2mm}
\epsfig{figure=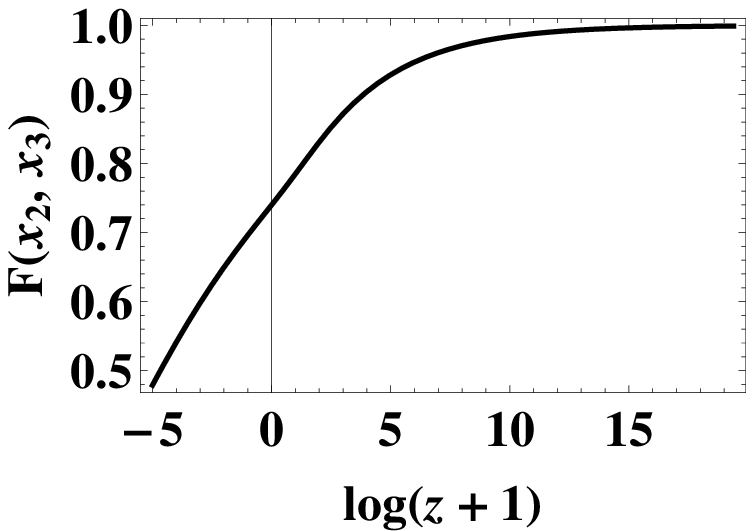,width=4.25cm}\hspace{2mm}
\epsfig{figure=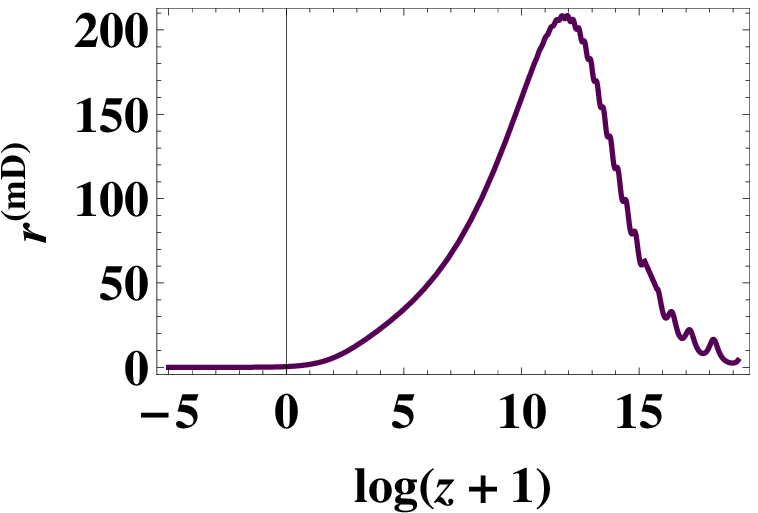,width=4.25cm}\vspace{2mm}
\epsfig{figure=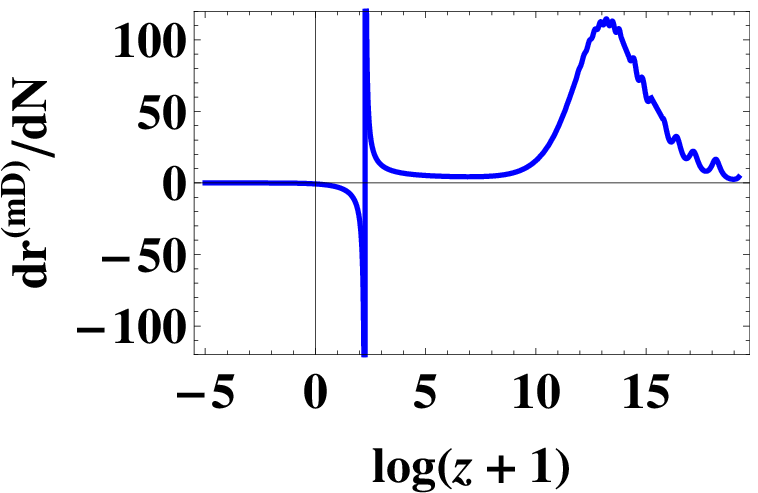,width=4.25cm}\hspace{2mm}
\epsfig{figure=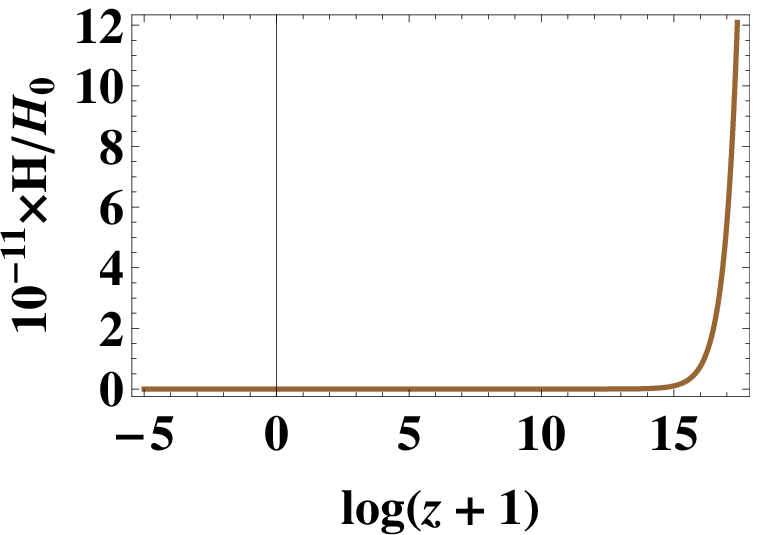,width=4.25cm}\hspace{2mm}
\epsfig{figure=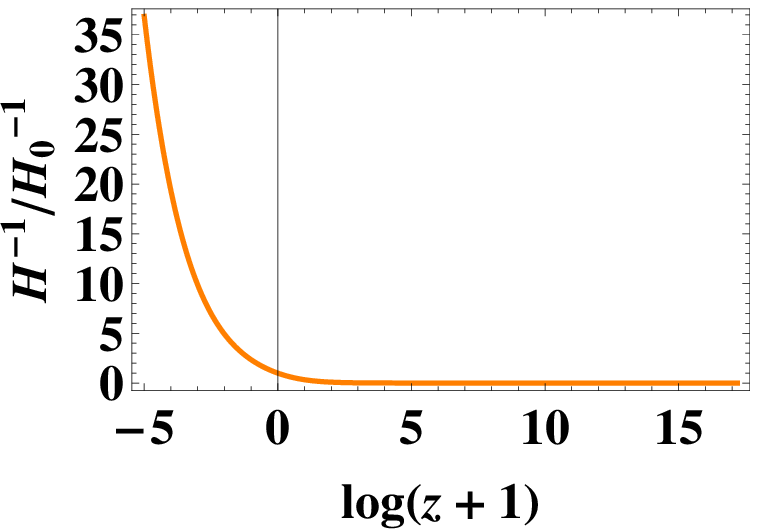,width=4.25cm}\hspace{2mm}
\epsfig{figure=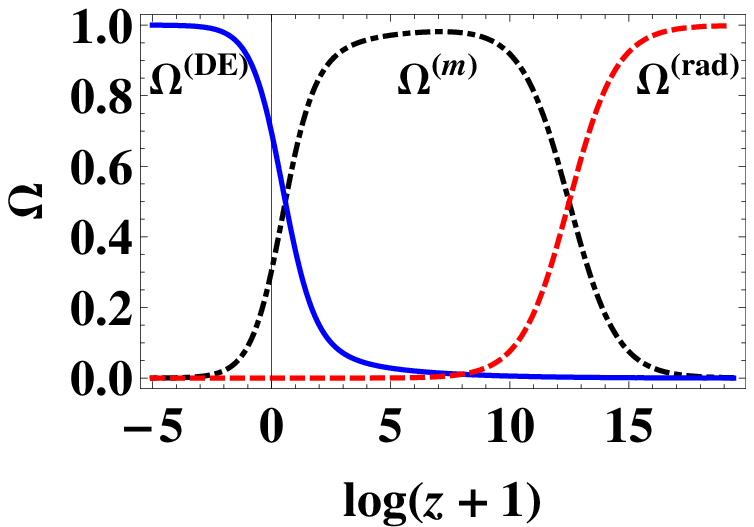,width=4.25cm}\vspace{2mm}
\epsfig{figure=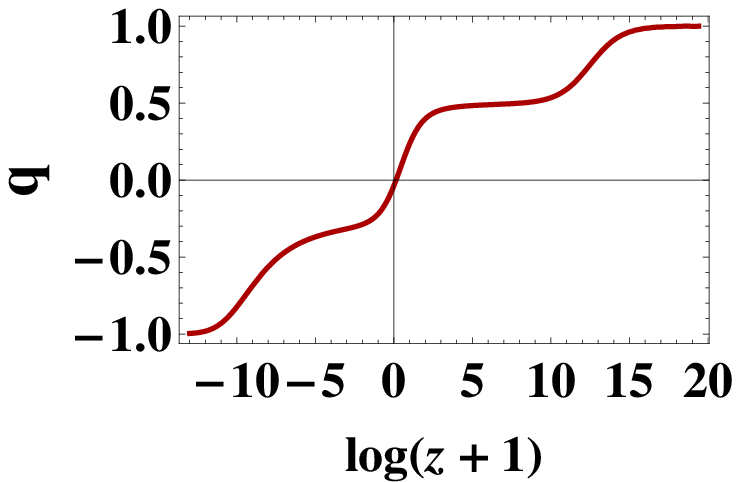,width=4.25cm}\hspace{2mm}
\epsfig{figure=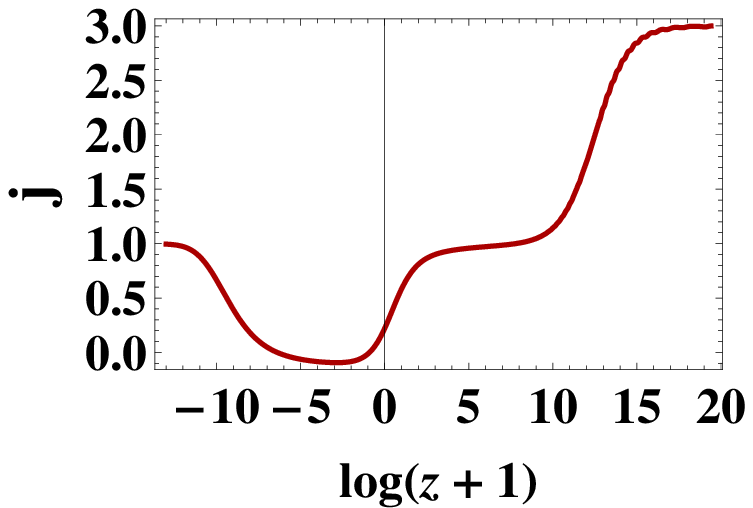,width=4.25cm}\hspace{2mm}
\epsfig{figure=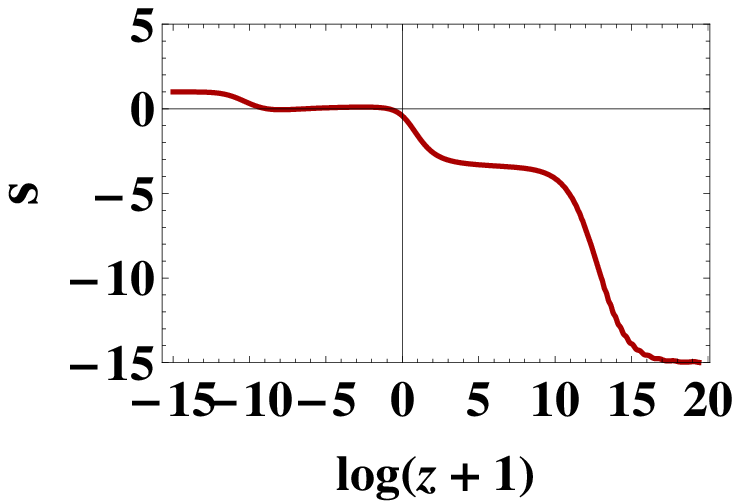,width=4.25cm}\hspace{2mm}
\caption{(color online).\footnotesize {\textbf{ Cosmological
solutions of $f(R,T)=R+ \alpha R^{0.9}+\sqrt{-T}$ gravity.}} For
this theory, we have chosen the initial values $x_{1}=10^{-6}$,
$x_{2}=-10^{-3}$, $x_{3}=1.0001 \times 10^{-3}$, $x_{4}=0.45
\times 10^{-15}$, and $x_{5}=0.999$ that correspond to
$z\thickapprox 2.6 \times 10^{8}$ with $\alpha = 4.06 \times
10^{-5}$. The initial values are chosen in a way that the matter
era occurs in the redshift of about $z_{m}\simeq 1100$, which
corresponds to the last scattering surface. With these initial
values, the radiation--matter equality takes place at
$z\simeq10^{5}$, which is far from the given results in the
literature. Note that it is possible to choose some other initial
values to solve this inconsistency, however, one should be careful
about the other cosmological quantities, such as the value of the
function $F$. The above diagrams for the deceleration, jerk, and
snap parameters show a consistent sequence for different
cosmological eras. However, their present values do~not match the
present observational data. For this theory, the relative size of
the universe in this era is about $1.2 \times 10^{-5}$.}
\label{CaseE2}
\end{figure}

In this case, from definitions (\ref{parameterm})
and (\ref{parameterr}), we have
\begin{align}\label{F(x)-B}
m=-n\frac{x_{3}+x_{2}}{x_{3}},~~~~~~~~~~~~~~~~~~~~~~~~~~\frac{dm}{dN}=x_{1}(1-\frac{nx_{2}}{x_{3}}),
\end{align}
\begin{align}\label{F(x)-B}
F(x_{2},x_{3})=(1+n)\frac{x_{3}}{x_{3}-nx_{2}}~~~~~~~~~~\mbox{and}~~~~~~~~~~
H(x_{2},x_{3})=\sqrt{\frac{[\frac{x_{2}+x_{3}}{\alpha(nx_{2}-x_{3})}]^{\frac{-1}{1+n}}}{6 x_{3}}},
\end{align}
where $r\neq n$. As we have indicated before, for
$-\sqrt{|n|}<r_{i}<-1$ with $n\rightarrow-1^{-}$, the effective
equation--of--state parameter has the value $-0.5$ (i.e., the
related fixed point in the phase space is stable), and otherwise
it gets $-1$. This means that it is important to choose an
appropriate initial value for $r_{i}$ for consistency with the
observations. Here, the behavior of the coincidence parameter, its
derivative, and the dark energy equation--of--state parameter, for
the model with $n=-1.01$, are shown in Figure~\ref{CaseE1}. On the
other hand, for the model with $n=-0.9$, we can set the initial
values in a way that the pick of the matter density diagram occurs
in the redshift about $ z\simeq 1100$. The price for this
consistency is that the matter--dominated era lasts for a long
time. The relative size of the universe in this era is about $1.2
\times 10^{-5}$, which is a more consistent value~\cite{mukhanov}.
The related diagram for this model is drawn in
Figure~\ref{CaseE2}. It is obvious that $F(x_{2},x_{3})$ has
approximately the value around one in the deep matter and
radiation eras (i.e., $F\simeq 0.97$), which means $g(R)\thicksim
R$ in such times. Again, the value $r^{\textrm{(mD)}}\simeq 3/7$
appears twice with $ dr^{\textrm{(mD)}}/dN \simeq -0.6$. We have
also plotted the related diagrams of the decelerated, the jerk,
and the snap parameters in Figure~\ref{CaseE2}. These diagrams
show a consistence sequence for different cosmological eras,
however, their present values $q_{0}\simeq -0.04$,
$j_{0}\simeq0.22$, and $s_{0}\simeq -0.76$ do~not match the
present observational data. The matter--dominated era for the
theory with $n=-1.01$ occurs in the redshift $z\sim 43$ with the
relative size of about $6 \times 10^{-3}$. We have summarized the
discussed features of these four models in Table~\ref{Tab2}, where
the values of the deceleration, jerk, and snap parameters for the
first three models are also presented.
\\\section{Cosmological Solution of Model $f(R,T)=R+c_{1}\sqrt{-T}$}\label{simplecase}
In this section, we investigate the cosmological properties of
theory $f(R,T)=R+c_{1}\sqrt{-T}$. This theory is similar to
theories with the dynamical cosmological constant. However, in
contrast to the most of them, this theory does~not include any
scalar field as the matter field, which motivates one to extract
the cosmological solutions of this theory. As in this case, we
have $m(r)=0$, and the formalism presented in the first section
does~not work, hence, we should obtain the dynamical system
equations independently. To complete the discussion, we include
the spatial curvature term to investigate whether $f(R,T)$ gravity
can select a particular sign for the curvature constant $k$. From
equations (\ref{first}) and (\ref{second}) for $g(R)=R$, one
obtains
\begin{align}\label{simple1}
1+c_{1}\frac{h}{6H^2}+\frac{k}{a^2 H^2}=\frac{8\pi G \rho^{\textrm{(m)}}}{3H^2}
+c_{1}\frac{h'\rho^{\textrm{(m)}}}{3H^2}+\frac{8\pi G \rho^{\textrm{(rad)}}}{3H^2}
\end{align}
and
\begin{align}\label{simple2}
\frac{2\dot{H}}{H^2}-\frac{2k}{a^2 H^2}=-\frac{8\pi G \rho^{\textrm{(m)}}}{H^2}-c_{1}\frac{h' \rho^{\textrm{(m)}}}{H^2}-\frac{32\pi G \rho^{\textrm{(rad)}}}{3H^2},
\end{align}
where $k$ is normalized to $+1$, $0$, and $-1$ for a closed, flat,
and open universe, respectively. Note that, the field equations
(\ref{simple1}) and (\ref{simple2}) do~not contain the variables
$x_{1}$, $x_{2}$, and $x_{5}$, and therefore the only contained
variables are $x_{3}$, $x_{4}$, $\Omega^{\textrm{(k)}}$,
$\Omega^{\textrm{(rad)}}$, and $\Omega^{\textrm{(DE)}}$. Equation
(\ref{simple1}) gives
\begin{align}\label{simple-omegas}
\Omega^{\textrm{(m)}}+\Omega^{\textrm{(k)}}+\Omega^{\textrm{(rad)}}+c_{1}\Omega^{\textrm{(DE)}}=1,
\end{align}
where from equation (\ref{Dark energy}), and the definitions
$x_{1}$ and $x_{2}$, in this case, we have
$\Omega^{\textrm{(DE)}}= x_{4}$. Therefore, we incorporate four
fluid densities, namely, the matter, spatial curvature, radiation,
and dark energy ones.

On the other hand, by using equations (\ref{varx3}) and
(\ref{simple2}), we reach at the following constraint
\begin{align}\label{simple-varx3}
x_{3}=\frac{1}{2}-\frac{\Omega^{\textrm{(k)}}}{2}-\frac{\Omega^{\textrm{(rad)}}}{2}+\frac{3c_{1}\Omega^{\textrm{(DE)}}}{4}.
\end{align}
Thus, these two constraints, equations (\ref{simple-omegas}) and
(\ref{simple-varx3}), eliminate the two variables $x_{3}$ and
$\Omega^{\textrm{(m)}}$, and only three dimensionless density
parameters, $\Omega^{\textrm{(k)}}$, $\Omega^{\textrm{(rad)}}$,
and $\Omega^{\textrm{(DE)}}$, remain as the independent variables.
For the effective equation--of--state parameter, we obtain
\begin{align}\label{simple-effective Eom}
w^{\textrm{(eff)}}=\frac{1}{3}\left(\Omega^{\textrm{(rad)}}
-\Omega^{\textrm{(k)}}\right)-c_{1}\frac{\Omega^{\textrm{(DE)}}}{2}.
\end{align}
Hence, we have a three--variable phase space with the following
equations of motion
\begin{align}
&\frac{d \Omega^{\textrm{(k)}}}{d N}=\Omega^{\textrm{(k)}}\left( 1+\Omega^{\textrm{(rad)}}-
\Omega^{\textrm{(k)}}-c_{1}\frac{3\Omega^{\textrm{(DE)}}}{2} \right)\label{simple-EOM1},\\
&\frac{d \Omega^{\textrm{(rad)}}}{d N}=\Omega^{\textrm{(rad)}}\left( -1+\Omega^{\textrm{(rad)}}-\Omega^{\textrm{(k)}}-
c_{1}\frac{3\Omega^{\textrm{(DE)}}}{2} \right)\label{simple-EOM2},\\
&\frac{d \Omega^{\textrm{(DE)}}}{d N}=\Omega^{\textrm{(DE)}}\left( \frac{3}{2}+\Omega^{\textrm{(rad)}}-\Omega^{\textrm{(k)}}-
c_{1}\frac{3\Omega^{\textrm{(DE)}}}{2} \right)\label{simple-EOM3}.
\end{align}

In Table~\ref{Tab3}, we have shown the fixed points and their
stability properties. There are four fixed points comprised of the
point $P^{\textrm{(DE)}}$ that determines an
acceleration--expansion--dominated era, the point
$P^{\textrm{(k)}}$ that refers to an era in which the spatial
curvature density is dominated over the other densities, the point
$P^{\textrm{(rad)}}$ that indicates a radiation--dominated era,
and, finally, the point $P^{\textrm{(m)}}$ that points to a
matter--dominated era. It is clear that only the dark energy fixed
point is stable, for the appearance of negative eigenvalues. It
means that the dark--energy--dominated era is an everlasting era.
Nevertheless, this fixed point corresponds to
$w^{\textrm{(eff)}}=-1/2$, which is~not an observational
consistent value. The saddle fixed point $P^{\textrm{(k)}}$
indicates $\Omega^{\textrm{(k)}}=1$, which refers to an open
universe. There is no solution with $\Omega^{\textrm{(k)}}=-1$ as
a closed universe.
\begin{center}
\begin{table}[h]
\centering
\caption{The fixed points solutions of $f(R,T)=R+c_{1}\sqrt{-T}$ gravity radiation.}
\begin{tabular}{l @{\hskip 0.1in} l@{\hskip 0.1in} l @{\hskip 0.1in}l @{\hskip 0.1in}l}\hline\hline

Fixed point     &Coordinates $(\Omega^{\textrm{(k)}},\Omega^{\textrm{(rad)}},\Omega^{\textrm{(DE)}})$           &Eigenvalues  &$\Omega^{\textrm{(m)}}$      &$w^{\textrm{(eff)}}$\\[0.5 ex]
\hline
$P^{\textrm{(DE)}}$&$(0,0,\frac{1}{c_{1}})$&$(-\frac{5}{2},-\frac{3}{2},-\frac{1}{2})$&$0$&$-\frac{1}{2}$\\[0.75 ex]
$P^{\textrm{(k)}}$&$(1,0,0)$&$(-2,-1,\frac{1}{2})$&$0$&$-\frac{1}{3}$\\[0.75 ex]
$P^{\textrm{(rad)}}$&$(0,1,0)$&$(5,2,1)$&$0$&$\frac{1}{3}$\\[0.75 ex]
$P^{\textrm{(m)}}$&$(0,0,0)$&$(\frac{3}{2},-1,1)$&$1$&$0$\\[0.75 ex]
\hline\hline
\end{tabular}
\label{Tab3}
\end{table}
\end{center}

\begin{figure}[h]
\epsfig{figure=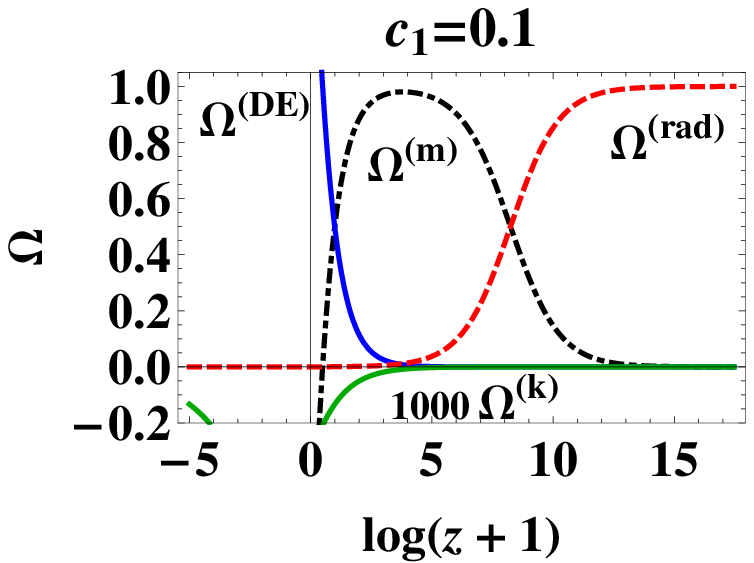,width=4.15cm}\hspace{2mm}
\epsfig{figure=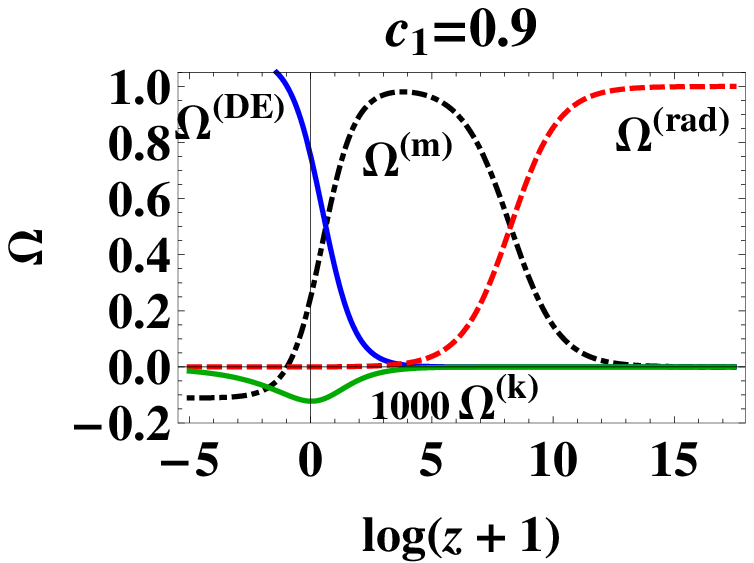,width=4.15cm}\vspace{2mm}
\epsfig{figure=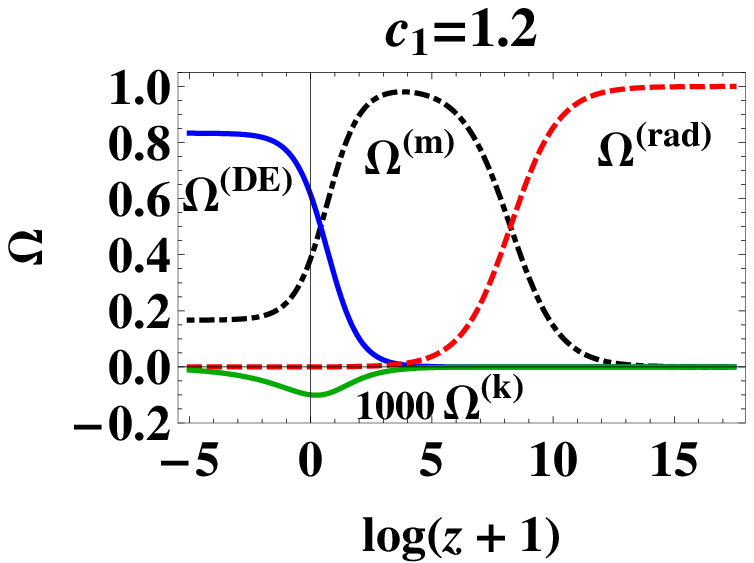,width=4.35cm}\hspace{2mm}
\epsfig{figure=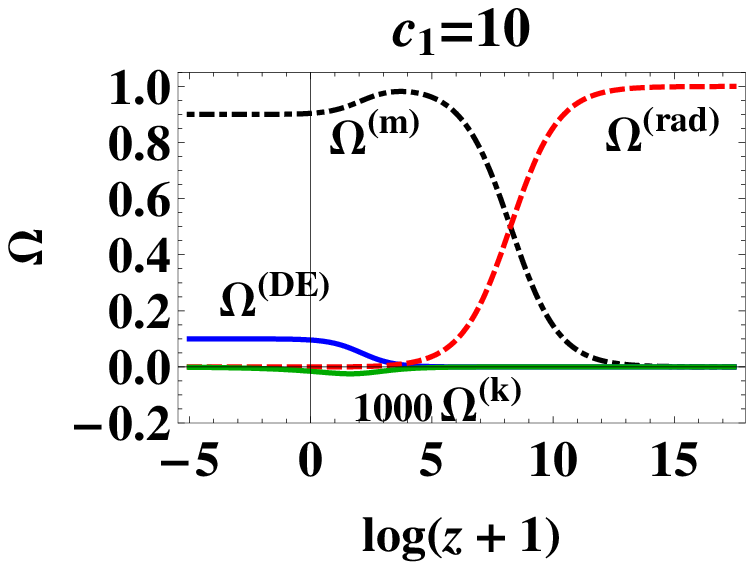,width=4.5cm}\ \caption{(color
online).\footnotesize{\textbf{ The effects of coupling constant
$c_{1}$ on the fluid density parameters in
$f(R,T)=R+c_{1}\sqrt{-T}$ gravity for an open universe.}} The
radiation density parameter is~not affected by this constant. The
larger (smaller) values of $c_{1}$ leads to small (larger) values
of the dark energy and the spatial curvature densities.}
\label{new1}
\end{figure}

As Table~\ref{Tab3} shows, the dark energy density parameter
depends on the coupling constant $c_{1}$, and hence it is
constrained to have positive values. However, it does~not affect
the stability properties of the fixed points and the value of the
equation--of--state parameter. In Figure~\ref{new1}, we have
depicted the density parameters for an open universe. These
diagrams show that for large values of $c_{1}$ the dark energy
density parameter tends to small values, which means that the
late--time era is effectively dominated by dust--like matter. In
such situations, the magnitude of the spatial curvature density
decreases to smaller values. On the other hand, for $0<c_{1}<1$,
the pressureless matter density gets negative values, which is~not
physically justified. Also in these cases, the magnitude of the
spatial curvature density increases to larger values. We have
checked that the results for a closed universe are the same as the
ones for an open universe. Nevertheless, to have a dominant dark
energy era with $\Omega^{\textrm{(DE)}}\simeq1$, the preferred
value for the coupling constant is $c_{1}\simeq1$ (in our assumed
units), and we will adopt this value in the rest of this work.
Thus, the terms $R$ and $\sqrt{-T}$ are placed of the same order
of magnitude in the Lagrangian (again in our assumed units). Also,
we have checked that the minimal models $f(R,T)=g(R)+h(T)$,
discussed in Sec.~\ref{Fieldequation}, show similar behaviors.

\begin{figure}[h]
\epsfig{figure=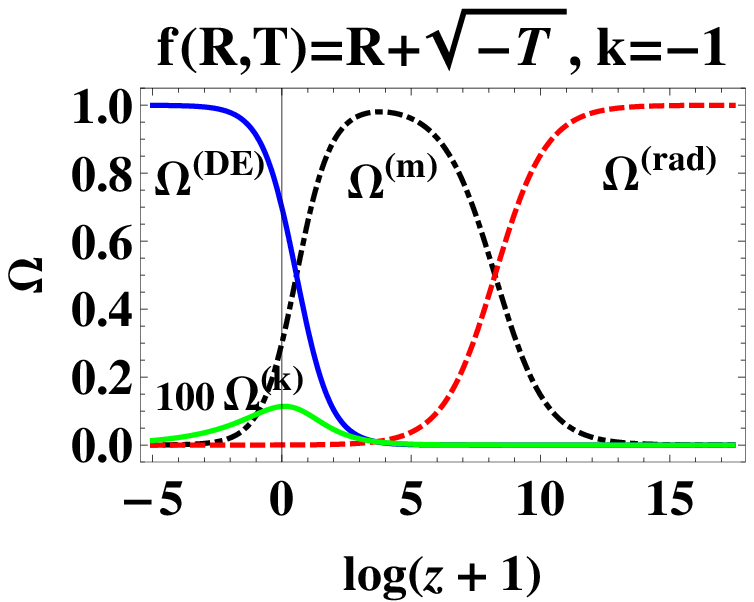,width=4.15cm}\hspace{2mm}
\epsfig{figure=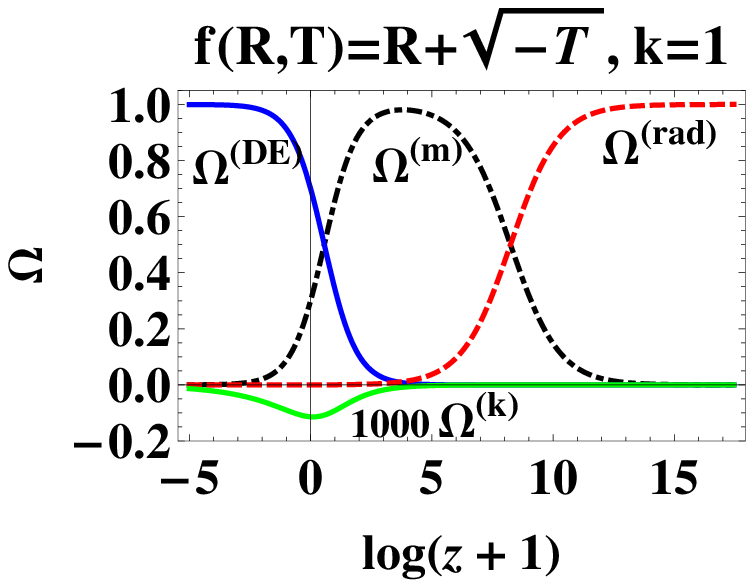,width=4.15cm}\vspace{2mm}
\epsfig{figure=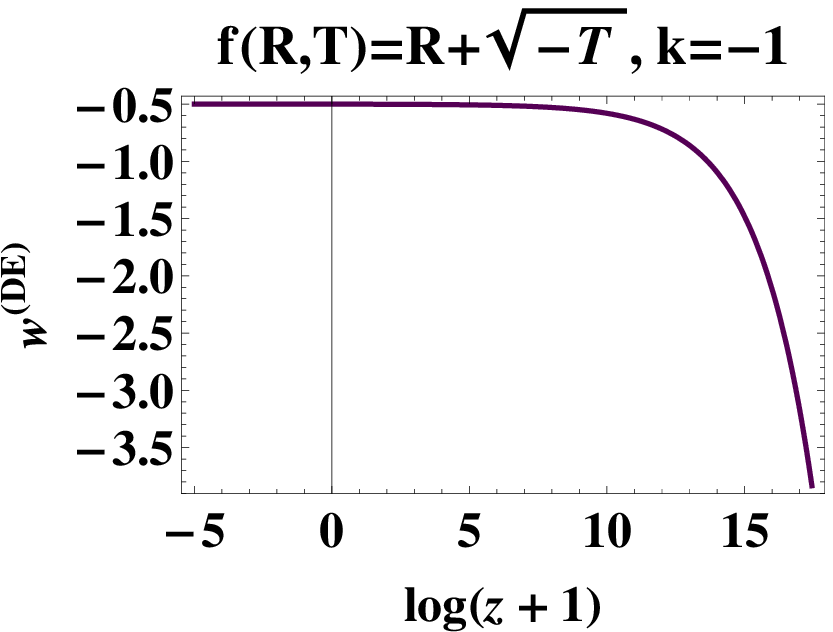,width=4.35cm}\hspace{2mm}
\epsfig{figure=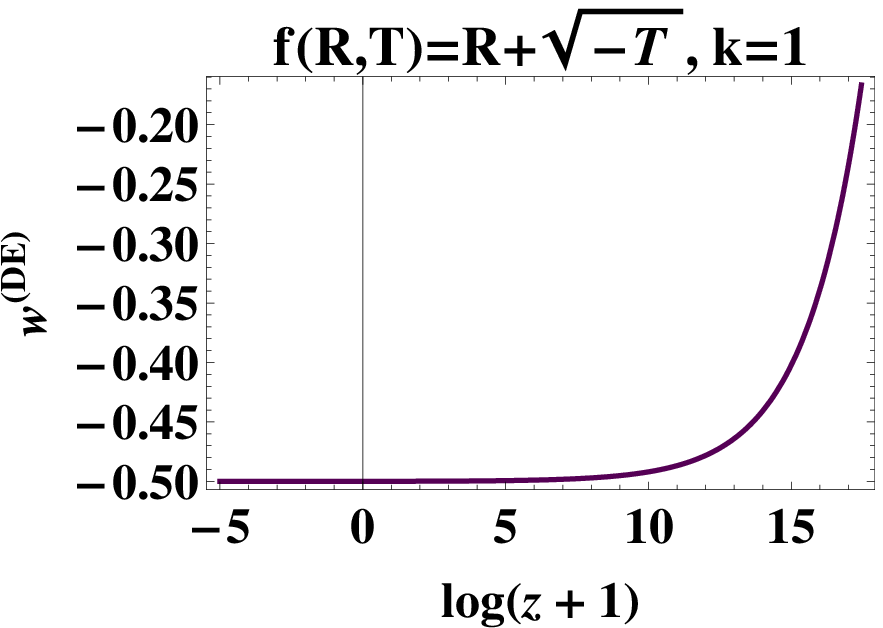,width=4.5cm}\ \caption{(color
online).\footnotesize{\textbf{ Cosmological solutions of
$f(R,T)=R+c_{1}\sqrt{-T}$ gravity with $c_{1}=1$.}} These diagrams
accept the initial values $\Omega^{\textrm{(rad)}}=0.9999$,
$\Omega^{\textrm{(DE)}}=10^{-15}$, and
$\Omega^{\textrm{(k)}}=10^{-15}$ for an open universe and the same
values, but with $\Omega^{\textrm{(k)}}=-10^{-15}$, for a closed
universe, corresponding to $z\thickapprox 3.6 \times 10^{7}$. The
diagrams show $\Omega^{\textrm{(k)}}_{0}=10^{-3}$ for the open
universe and $\Omega^ {\textrm{(k)}}_{0}=-10^{-4}$ for the closed
universe. The dark energy curve has an increasing behavior up to
the value $w^{\textrm{(eff)}}=-1/2$ for the open universe, and it
has a decreasing behavior down to the value
$w^{\textrm{(eff)}}=-1/2$ for the closed universe.}
\label{Simple1}
\end{figure}

\begin{figure}[h]
\epsfig{figure=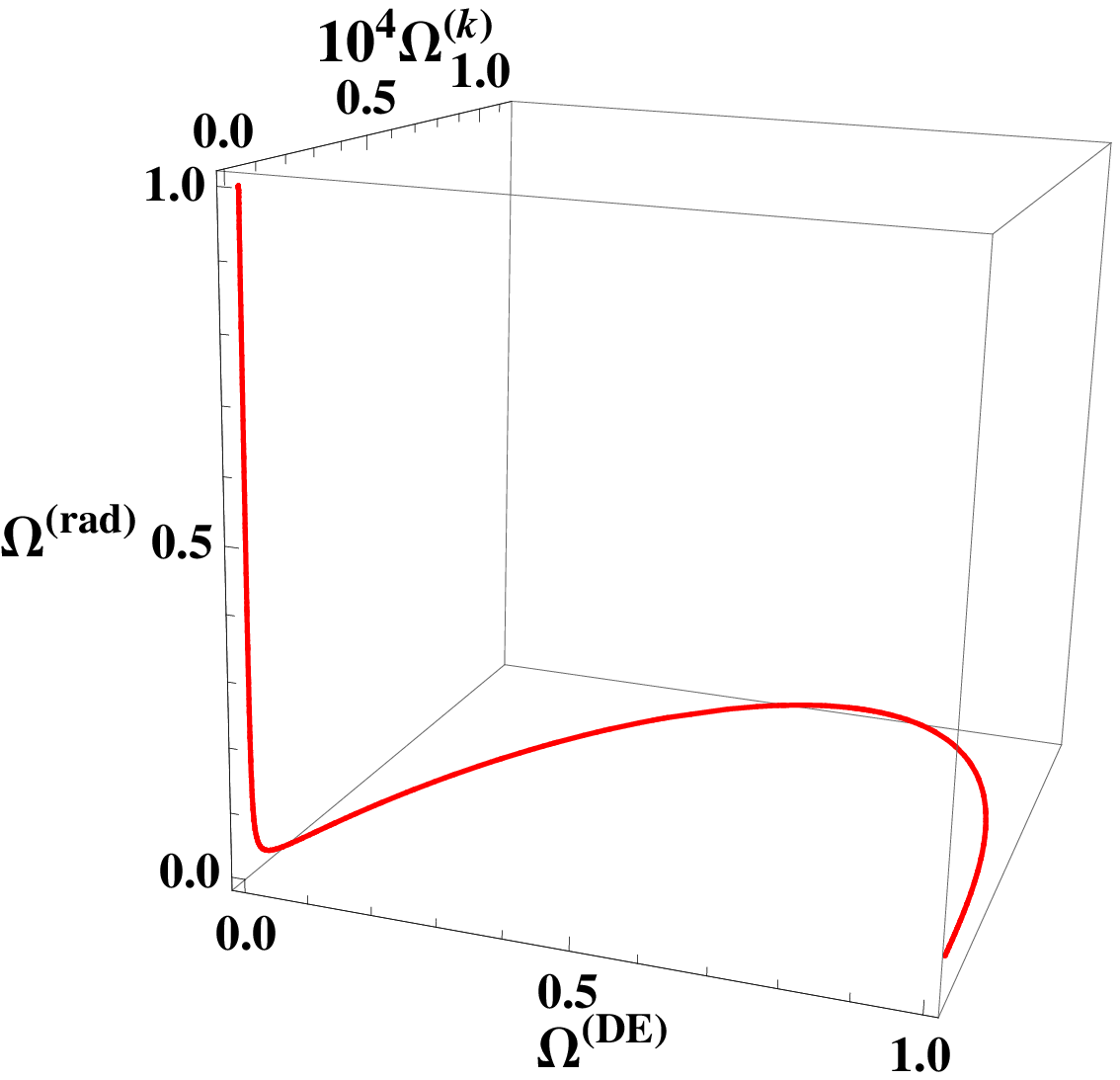,width=6cm}\hspace{30mm}
\epsfig{figure=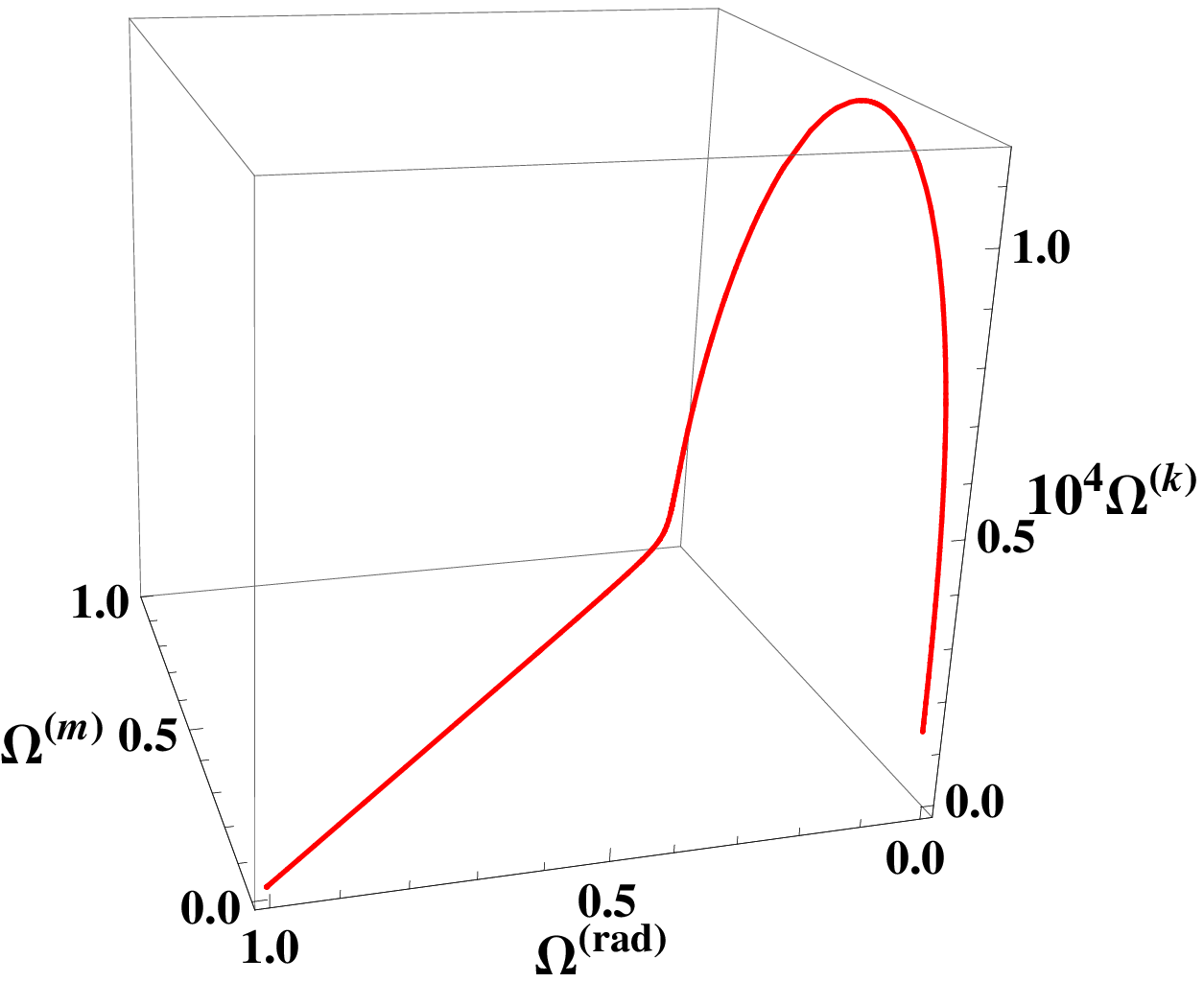,width=6.5cm} \caption{(color
online).\footnotesize{\textbf{ Cosmological solutions of
$f(R,T)=R+c_{1}\sqrt{-T}$ gravity for $k=-1$.}} The left
parametric plot is in terms of $\Omega^{\textrm{(k)}}$,
$\Omega^{\textrm{(rad)}}$, and $\Omega^{\textrm{(DE)}}$, and the
right one is in terms of $\Omega^{\textrm{(k)}}$, $\Omega^
{\textrm{(rad)}}$, and $\Omega^{\textrm{(m)}}$ coordinates. The
peaks of the curves occur at $\Omega^{\textrm{(DE)}}\simeq0.7$ in
the left and at $\Omega^{\textrm{(m)}}\simeq0.3$ in the right
diagrams, respectively.} \label{Simple2}
\end{figure}

\begin{figure}[h]
\epsfig{figure=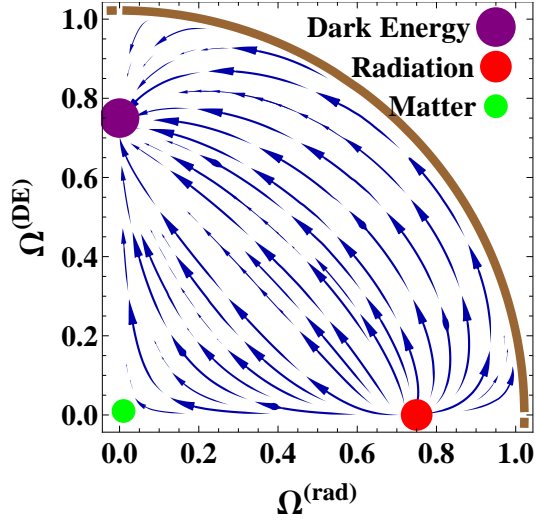,width=7cm} \caption{(color online). A
phase portrait for the special case $k=0$ is drawn. There is an
acceptable trajectory that connects the radiation era to the
matter era then to the dark--energy--dominated era as the final
attractor. The other trajectories do~not contain any proper
cosmological sequence.} \label{Simple3}
\end{figure}

We have drawn the diagrams for the density components and the dark
energy equation of state for the both positive and negative
initial values of $\Omega^{\textrm{(k)}}$ in Figure~\ref{Simple1}.
As is obvious, for the both initial values, the amplitude of
$\Omega^{\textrm{(k)}}$ is~not far from the observational data,
i.e., the diagrams also indicate that for an open universe we have
$\Omega^{\textrm{(k)}}\simeq 10^{-3}$ and for a closed one
$\Omega^{\textrm{(k)}} \simeq 10^{-4}$. In Figure~\ref{Simple1},
the diagrams show that $\Omega^{\textrm{(k)}}$, in the
matter--dominated era, is small and the curvature domination
occurs in the late times and large scales. The other diagrams
illustrate the evolution of the equation--of--state parameter for
the dark energy, both in an open and a closed universe. For an
open (closed) universe, the corresponding curve has an increasing
(decreasing) feature to the value $w^{\textrm{(eff)}}=-1/2$. This
contrast gives us an opportunity to match it with the
observations. Irrespective of the non--consistent value
$w^{\textrm{(eff)}}=-1/2$, the increasing (decreasing) feature can
be a distinguishable criteria. If the evolution of
$w^{\textrm{(eff)}}$ in different cosmological eras can be
investigated from the observational data, then the increasing
(decreasing) feature will rule out any inconsistent cosmological
model.
In Figure~\ref{Simple2}, we have presented a parametric plot in
$(\Omega^{\textrm{(k)}}, \Omega^{\textrm{(rad)}},
\Omega^{\textrm{(DE)}})$ coordinates for the left diagram and in
$(\Omega^{\textrm{(k)}},
\Omega^{\textrm{(rad)}},\Omega^{\textrm{(m)}})$ coordinates for
the right one, both in an open universe. Both diagrams show an
acceptable evolution of the density parameters. The left figure
illustrates that $\Omega^{\textrm{(k)}}$ has a peak when
$\Omega^{\textrm{(DE)}}\simeq0.7$, and the right figure indicates
that it happens when $\Omega^{\textrm{(m)}}\simeq0.3$.
Also, in Figure~\ref{Simple3}, we have depicted a phase portrait
of the solutions in the surface $\Omega^{\textrm{(k)}}=0$. In this
plot, for some initial values, we have a few trajectories that,
after leaving the radiation fixed point, approach the matter fixed
point and finally are attracted to the dark energy fixed point.
Among the trajectories, there are ones that directly connect the
early era (corresponding to the radiation fixed point) to the
late--time era (corresponding to the dark energy fixed point)
without passing the matter era (corresponding to the matter fixed
point). Some trajectories connect the matter era to the
accelerated expansion era without starting from the radiation era,
etc. Note that the negative valued regions are~not related to our
solutions.
\section{Weak--field limit of $f(R,T)$ gravity models}\label{weakfield}
In this section, we consider the weak--field limit of $f(R,T)$
gravity\rlap.\footnote{The Newtonian limit of $f(R,T)$ gravity has
also been considered in Ref.~\cite{harko} via another approach
(wherein the authors have assumed that the matter component is~not
conserved) and hence, have obtained a different result for the
weak--field limit of $f(R,T)$ gravity. Incidentally, a modified
version of their work is presented in Ref.~\cite{correct}.} To
perform this task, we implicitly assume those models that admit
Taylor series expansion around the background values of the Ricci
scalar and the trace of the energy--momentum tensor. In addition,
we work on a pressureless matter, where in this case, the
magnitude of the trace is the same as the mass density of matter.
To consider the weak--field limit of $f(R,T)$ gravity, we first
linearize the field equation (\ref{field tensor}) (which we
preferably rewrite as
\begin{align}\label{field-back1}
F(R,T) R_{\mu \nu}-\frac{1}{2} f(R,T) g_{\mu \nu}+\Big{(} g_{\mu \nu}
\square -\triangledown_{\mu} \triangledown_{\nu}\Big{)}F(R,T)=\Big{(}8
\pi G+ {\mathcal F}(R,T)\Big{)} T_{\mu \nu},
\end{align}
where $T_{\mu \nu}$ generally depends on the spatial coordinates
and time). In comparison, in $f(R)$ gravity, only the matter
density perturbation sources the perturbation of the Ricci scalar,
but in $f(R,T)$ gravity, the perturbation of the trace of the
energy--momentum tensor also sources the perturbation of the
curvature of space--time. Here, we investigate the space--time
metric outside a spherical body and then determine the PPN gamma
parameter to confront the $f(R,T)$ models with the observational
data. We should remind the reader that, in the literature, the PPN
parameter of $f(R)$ gravity, without resorting to the
scalar--tensor equivalence, has been obtained to be
$\gamma^{\textrm{(f(R))}}= 1/2$~\cite{solar1}, nevertheless, in
this theory, using a scalar--tensor representation, one can still
acquire an observational consistent value\rlap,\footnote{We have
provided a very concise discussion on the corresponding issue of
this scalar--tensor representation in the following part (just
before considering the General Minimal Power Law Case).}\
 which is supposed to be $\gamma^{\textrm{(obs)}}=1+(2.1\pm2.3) \times
10^{-5}$~\cite{ppn1,ppn2}.

Now, we suppose a background space--time that is usually assumed
to be the isotropic and homogeneous one. This space--time admits a
background spatially uniform energy--momentum tensor\footnote{As
the radiation matter does~not have any effect on the solar system
experiments, it has~not been included in this section.} $T_{\mu
\nu}^{\textrm{(b)}}$. For the background space--time with a
background energy--momentum tensor $T_{\mu \nu}^{\textrm{(b)}}$,
we have
\begin{align}\label{field-back2}
F^{\textrm{(b)}}(t) R_{\mu \nu}^{\textrm{(b)}}(t)-\frac{1}{2} f^{\textrm{(b)}}(t)
g_{\mu \nu}^{\textrm{(b)}}(t)+\Big{(} g_{\mu \nu}^{\textrm{(b)}}(t)
\square -\triangledown_{\mu} \triangledown_{\nu}\Big{)}F^{\textrm{(b)}}(t)=\Big{(}8
\pi G+ \mathcal{F}^{\textrm{(b)}}(t)\Big{)} T^{\textrm{(b)}}_{\mu \nu}(t),
\end{align}
where $F^{\textrm{(b)}}\equiv\partial f/\partial
R|_{(R^{\textrm{(b)}},T^{\textrm{(b)}})}$, $f^{\textrm{(b)}}\equiv
f(R^{\textrm{(b)}},T^{\textrm{(b)}})$, and
$\mathcal{F}^{\textrm{(b)}}\equiv
\partial f/\partial T|_{(R^{\textrm{(b)}},T^{\textrm{(b)}})}$.
The supper script $b$ denotes the background quantities, and we
have dropped the argument $(R,T)$. By the argument $t$, we have
indicated that all the background quantities depend only on the
cosmic time. Using the trace of equation (\ref{field-back1}), one
can determine the evolution of time--dependent scalar curvature
$R^{\textrm{(b)}}(t)$ as
\begin{align}\label{trace-back}
F^{\textrm{(b)}}(t)R^{\textrm{(b)}}(t)+3\Box F^{\textrm{(b)}}(t)-2f^{\textrm{(b)}}(t)=
(8\pi G+\mathcal{F}^{\textrm{(b)}}(t))T^{\textrm{(b)}}(t).
\end{align}

To linearize the field equation
(\ref{field-back1}) and its trace equation, we perturb the Ricci scalar,
the Ricci tensor, the energy--momentum tensor and its trace as the sum of a time--dependent spatially
homogeneous background component and a time--independent source component, namely
\begin{align}\label{perturbation}
&R(t,r)=R^{\textrm{(b)}}(t)+R^{\textrm{(s)}}(r),\\
&R_{\mu \nu}(t,r)=R^{\textrm{(b)}}_{\mu \nu}(t)+R^{\textrm{(s)}}_{\mu \nu}(r),\\
&T_{\mu \nu}(t,r)=T_{\mu \nu}^{\textrm{(b)}}(t)+T_{\mu \nu}^{\textrm{(s)}}(r),\\
&T(t,r)=T^{\textrm{(b)}}(t)+T^{\textrm{(s)}}(r).
\end{align}
The background metric is taken to be the FLRW metric
(\ref{metricFRW}) with $k=0$, and its spherically symmetric
linearized form reads
\begin{align}\label{metric-pert}
ds^{2}=-\Big{(}1+2\Psi(r)\Big{)}dt^{2}+a^{2}(t)\left[\Big{(}1+2\Phi(r)\Big{)}dr^{2}+r^{2}d\Omega^{2}\right],
\end{align}
where $\Psi(r)$ and $\Phi(r)$ are the metric perturbations. To
proceed, we use the Taylor expansion of the functions $f(R,T)$,
$F(R,T)$, and $\mathcal{F}(R,T)$ around the background quantities
$R^{\textrm{(b)}}$ and $T^{\textrm{(b)}}$. Further, we neglect the
non--linear terms in the expansions, assuming that the zeroth and
the first terms dominate these terms. Therefore, the first--order
version of the trace of equation (\ref{field-back1}) is obtained
as
\begin{align}\label{R1-equation-1}
\nabla^{2}R^{\textrm{(s)}}-m^{2}_{\textrm{f(R,T)}}R^{\textrm{(s)}}=\mathcal{S},
\end{align}
where we have defined a mass parameter
\begin{align}\label{mass1}
m^{2}_{\textrm{f(R,T)}}\equiv \frac{1}{3}\left(\frac{F^{\textrm{(b)}}}{F_{R}^{\textrm{(b)}}}+
\frac{\mathcal{F}_{\textrm{R}}^{\textrm{(b)}}}{F_{\textrm{R}}^{\textrm{(b)}}}T^{\textrm{(b)}}-R^{\textrm{(b)}}+
\frac{3}{F_{\textrm{R}}^{\textrm{(b)}}}\frac{d^{2}F_{\textrm{R}}^{\textrm{(b)}}}{dt^{2}}\right),
\end{align}
a source parameter
\begin{align}\label{source1}
\mathcal{S}\equiv \frac{1}{3F_{\textrm{R}}^{\textrm{(b)}}}\left[\left(8\pi G+3\mathcal{F}^{\textrm{(b)}}
+\mathcal{F}_{\textrm{T}}^{\textrm{(b)}}T^{\textrm{(b)}}-F_{\textrm{T}}^{\textrm{(b)}}R^{\textrm{(b)}}
-3\frac{d^{2}F_{\textrm{T}}^{\textrm{(b)}}}{dt^{2}}\right)T^{\textrm{(s)}}-3F_{\textrm{T}}^{\textrm{(b)}}
\nabla^{2}T^{\textrm{(s)}}\right]
\end{align}
and we have used
$\Box\,\Xi^{\textrm{(b)}}(t)=-d^{2}\Xi^{\textrm{(b)}}(t)/dt^{2}$
and $\Box\,\Xi^{\textrm{(s)}}(r)=\nabla^{2}\Xi^{\textrm{(s)}}(r)$
for any arbitrary function $\Xi$. Also, we have defined
$F_{\textrm{R}}\equiv \partial F/\partial R$,
$F_{\textrm{T}}\equiv \partial F/\partial T$, $\mathcal{F}_
{\textrm{R}}\equiv \partial \mathcal{F}/\partial R$, and
$\mathcal{F}_{\textrm{T}} \equiv \partial\mathcal{F}/ \partial T$,
and in obtaining equation (\ref{R1-equation-1}), we have neglected
all the second--order terms like
$F_{\textrm{T}}^{\textrm{(b)}}T^{\textrm{(s)}}R^{\textrm{(s)}}$.
It is obvious that the mass parameter (\ref{mass1}) varies in
time, however, its variation is negligible with respect to the
cosmological time--scale variations that are of the order of the
current Hubble time. Hence, we take $m^{2}_{\textrm{f(R,T)}}$ to
be a time--independent parameter. Since, the time scale of the
solar system experiments is negligible compared to the
cosmological time scale, one can discard the time derivative of
all the quantities and therefore can use the present value of the
background quantities for the solar system applications. On the
other hand, $H^{2}$ is of the order of the inverse squared age of
the universe, and hence it is negligible at the present era,
similarly for the quantities $R^{\textrm{(b)}}\sim H^{2}$,
$d^{2}\ln{(F_{\textrm{R}}^{\textrm{(b)}})} /dt^{2}\sim
H^{2}$~\cite{solar1}. We also set the present value of the scale
factor as $1$. Therefore, the mass parameter (\ref{mass1}) and the
source expression (\ref{source1}) can be written as
\begin{align}\label{mass}
m^{2}_{\textrm{f(R,T)}}\simeq \frac{1}{3}\left(\frac{F^{\textrm{(b)}}}{F_{\textrm{R}}^{\textrm{(b)}}}+
\frac{\mathcal{F}_{\textrm{R}}^{\textrm{(b)}}}{F_{\textrm{R}}^{\textrm{(b)}}}T^{\textrm{(b)}}\right)
\end{align}
and
\begin{align}\label{source2}
\mathcal{S}\simeq \frac{1}{3F_{\textrm{R}}^{\textrm{(b)}}}\left[\left(8\pi G+3\mathcal{F}^{\textrm{(b)}}
+\mathcal{F}_{\textrm{T}}^{\textrm{(b)}}T^{\textrm{(b)}}\right)T^{\textrm{(s)}}-3F_{\textrm{T}}^{\textrm{(b)}}
\nabla^{2}T^{\textrm{(s)}}\right].
\end{align}

The study of the Green functions for equation
(\ref{R1-equation-1}) shows that in the limit
$|m^{2}_{\textrm{f(R,T)}}|r^{2}\ll1$, these functions
approximately take the form $-1/(4\pi r)$, which is the Green
function for the Laplace equation~\cite{solar1}. Hence, in this
limit, the second term in equation (\ref{R1-equation-1}) may be
neglected, and then the corresponding differential equation for
$R^{\textrm{(s)}}$ takes the following form
\begin{align}\label{R1-equation-2}
\nabla^{2}R^{\textrm{(s)}}\simeq\mathcal{S}.
\end{align}
In this case, the exterior solution of equation
(\ref{R1-equation-2}) for a spherical body with mass $M$, radius
$\Re$, and mass density $\rho^{\textrm{(s)}}(r)$ can be obtained
as
\begin{align}\label{solution-R1}
R^{\textrm{(s)}}\simeq \frac{2G^{\textrm{(eff)}}M}{3F_{\textrm{R}}^{\textrm{(b)}}r}+
\frac{F_{\textrm{T}}^{\textrm{(b)}}\mathcal{M}^{\textrm{(sur)}}(\Re)}{F_{\textrm{R}}^{\textrm{(b)}}r},
\end{align}
where we have assumed that all the perturbations vanish at long
distances. In solution (\ref{solution-R1}), we have defined a
quantity that has the dimension of mass per area of body surface,
i.e.,
\begin{align}\label{surface mass}
\mathcal{M}^{\textrm{(sur)}}(r)\equiv 4\pi r^{2}\frac{d\rho^{\textrm{(s)}}(r)}{dr}.
\end{align}
This term comes from $\nabla^{2}T^{\textrm{(s)}}$, which is a new
one in $f(R,T)$ gravity with respect to the $f(R)$ gravity worked
in Ref.~\cite{solar1}. Also, in solution (\ref{solution-R1}), we
have defined an effective gravitational constant
$G^{\textrm{(eff)}}$ as
\begin{align}\label{effective-G}
G^{\textrm{(eff)}}\equiv G+\frac{1}{8\pi }\left[3\mathcal{F}^{\textrm{(b)}}
+\mathcal{F}_{\textrm{T}}^{\textrm{(b)}}T^{\textrm{(b)}}\right].
\end{align}
Note that the condition $|m^{2}_{\textrm{f(R,T)}}|r^{2}\ll1$ leads
to
\begin{align}\label{constraint1}
\left|\frac{F^{\textrm{(b)}}}{F_{\textrm{R}}^{\textrm{(b)}}}+
\frac{\mathcal{F}_{\textrm{R}}^{\textrm{(b)}}}{F_{\textrm{R}}^{\textrm{(b)}}}T^{\textrm{(b)}}\right|r^{2}
=\left|\frac{F^{\textrm{(b)}}}{F_{\textrm{R}}^{\textrm{(b)}}}\left(1+\frac{\mathcal{F}_{\textrm{R}}^{\textrm{(b)}}}
{F^{\textrm{(b)}}}T^{\textrm{(b)}}\right)\right|r^{2},
\end{align}
where, provided that\footnote{This term vanishes for the minimal
models, however, one must check it for non--minimal $f(R,T)$
models.} $|(\mathcal{F}_{\textrm{R}}^{\textrm{(b)}}/
F^{\textrm{(b)}})T^{\textrm{(b)}}|\ll1$, the corresponding
condition for $f(R)$ gravity~\cite{solar1} is recovered, i.e.,
\begin{align}\label{constraint2}
\left|\frac{F^{\textrm{(b)}}}{F_{\textrm{R}}^{\textrm{(b)}}}\right|r^{2}\ll1.
\end{align}

Now, let us obtain the metric perturbations $\Psi(r)$ and $\Phi(r)$ by using
solution (\ref{solution-R1}). For this purpose, the first--order Taylor expansion of the field
equation (\ref{field-back1}) yields
\begin{align}\label{perturbed-field}
&F^{\textrm{(b)}}R^{\textrm{(s)}\nu}_{\mu}+\left(F_{\textrm{R}}^{\textrm{(b)}}R^{\textrm{(b)}\nu}
_{\mu}-\frac{1}{2}\delta^{\nu}_{\mu}F^{\textrm{(b)}}+\delta^{\nu}_{\mu}
F_{\textrm{R}}^{\textrm{(b)}}\Box-F_{\textrm{R}}^{\textrm{(b)}}
\nabla_{\mu}\nabla^{\nu}-\mathcal{F}_{\textrm{R}}^{\textrm{(b)}}T^{\textrm{(b)}\nu}_{\mu}+
\delta^{\nu}_{\mu}\Box F_{\textrm{R}}^{\textrm{(b)}}-\nabla_{\mu}\nabla^{\nu}
F_{\textrm{R}}^{\textrm{(b)}}\right)R^{\textrm{(s)}}=\nonumber\\
&\left(8\pi G+\mathcal{F}^{\textrm{(b)}}\right)T^{\textrm{(s)}\nu}_{\mu}+
\left(\mathcal{F}_{\textrm{T}}^{\textrm{(b)}}T^{\textrm{(b)}\nu}_{\mu}-
F_{\textrm{T}}^{\textrm{(b)}}R^{\textrm{(b)}\nu}_{\mu}+\frac{1}{2}\delta^{\nu}_{\mu}
\mathcal{F}^{\textrm{(b)}}-\delta^{\nu}_{\mu}F_{\textrm{T}}^{\textrm{(b)}}\Box+
F_{\textrm{T}}^{\textrm{(b)}}\nabla_{\mu}\nabla^{\nu}-\delta^{\nu}_{\mu}\Box
F_{\textrm{T}}^{\textrm{(b)}}+\nabla_{\mu}\nabla^{\nu}F_{\textrm{T}}^{\textrm{(b)}}
\right)T^{\textrm{(s)}},
\end{align}
of which the $tt$, $rr$, and $\theta\theta$ equations, for a
pressureless fluid, are approximated to give
\begin{align}
&F^{\textrm{(b)}}\nabla^{2}\Psi+\left(\frac{1}{2}F^{\textrm{(b)}}-\mathcal{F}_{\textrm{R}}^{\textrm{(b)}}
\rho^{\textrm{(b)}}\right)R^{\textrm{(s)}}-F_{\textrm{R}}^{\textrm{(b)}}\nabla^{2}R^{\textrm{(s)}}=
\left(8\pi G^{\textrm{(eff)}}-\frac{3}{2}\mathcal{F}^{\textrm{(b)}}+
F_{\textrm{T}}^{\textrm{(b)}}\nabla^{2}\right)\rho^{\textrm{(s)}},\label{perturbed-field1}\\
&F^{\textrm{(b)}}(-\Psi''+\frac{2}{r}\Phi')-\frac{1}{2}F^{\textrm{(b)}}R^{\textrm{(s)}}
+\frac{2}{r}F_{\textrm{R}}^{\textrm{(b)}}R'^{\textrm{(s)}}=-\frac{1}{2}\mathcal{F}^{\textrm{(b)}}\rho^{\textrm{(s)}}
+\frac{2}{r}F_{\textrm{T}}^{\textrm{(b)}}\rho'^{\textrm{(s)}},\label{perturbed-field2}\\
&F^{\textrm{(b)}}(\frac{1}{r}\Phi'-\frac{1}{r}\Psi'+\frac{2}{r^{2}}\Phi)
-\frac{1}{2}F^{\textrm{(b)}}R^{\textrm{(s)}}+\frac{1}{r}F_{\textrm{R}}^{\textrm{(b)}}R'^{\textrm{(s)}}
+F_{\textrm{R}}^{\textrm{(b)}}R''^{\textrm{(s)}}=-\frac{1}{2}\mathcal{F}^{\textrm{(b)}}\rho^{\textrm{(s)}}
+\frac{1}{r}F_{\textrm{T}}^{\textrm{(b)}}\rho'^{\textrm{(s)}}+F_{\textrm{T}}^{\textrm{(b)}}
\rho''^{\textrm{(s)}},\label{perturbed-field3}
\end{align}
where the prime indicates differentiation with respect to $r$. By
considering solution (\ref{solution-R1}), one does expect that
$\Psi(r)$ and $\Phi(r)$ will have similar functionality to $r$,
which implies that the terms $R^{\textrm{(s)}}\Psi$ and
$R^{\textrm{(s)}}\Phi$ are of second--order, and hence we have
neglected them in obtaining the filed equations
(\ref{perturbed-field1})-- (\ref{perturbed-field3}). Therefore,
the most general form of the potential $\Psi(r)$, outside the
body, using (\ref{R1-equation-2}), (\ref{solution-R1}), and
(\ref{perturbed-field1}), can be obtained as
\begin{align}\label{psi-gen}
\Psi(r)=-\frac{4G^{\textrm{(eff)}}M}{3F^{\textrm{(b)}}r}+\frac{3\mathcal{F}^{\textrm{(b)}}M}{8\pi
F^{\textrm{(b)}}r}-\left(\frac{G^{\textrm{(eff)}}M}{3F_{\textrm{R}}^{\textrm{(b)}}}+
\frac{F_{\textrm{T}}^{\textrm{(b)}}\mathcal{M}^{\textrm{(sur)}}(\Re)}{2F_{\textrm{R}}^{\textrm{(b)}}}\right)
\left(\frac{1}{2}-\frac{\mathcal{F}_{\textrm{R}}^{\textrm{(b)}}T^{\textrm{(b)}}}{F^{\textrm{(b)}}}\right)r
-\frac{C_{1\Psi}}{r}+C_{2\Psi},
\end{align}
where $C_{1\Psi}$ and $C_{2\Psi}$ are the integral constants and
we have assumed $F^{\textrm{(b)}}\neq 0$. One can set
$C_{2\Psi}=0$ as usually done in the Newtonian limit. In addition,
the term containing $C_{1\Psi}$ leads to singularity in the
origin, and hence one can discard it~\cite{solar1,solar2,solar3}.
Also, for the cases in which the condition
$|(\mathcal{F}_{\textrm{R}}^{\textrm{(b)}}/F^{\textrm{(b)}}
)T^{\textrm{(b)}}|\ll1$ holds, we rewrite $\Psi(r)$ as
\begin{align}\label{psi-1}
\Psi(r)\simeq-\frac{4G^{\textrm{(eff)}}M}{3F^{\textrm{(b)}}r}+\frac{3\mathcal{F}^{\textrm{(b)}}M}{8\pi
F^{\textrm{(b)}}r}-\frac{G^{\textrm{(eff)}}M}{6F_{\textrm{R}}^{\textrm{(b)}}}r+
\frac{F_{\textrm{T}}^{\textrm{(b)}}\mathcal{M}^{\textrm{(sur)}}(\Re)}{4F_{\textrm{R}}^{\textrm{(b)}}}
r.
\end{align}

We can further simplify solution (\ref{psi-1}) by comparing the
third and fourth terms with respect to the first term, i.e.,
\begin{align}\label{aprox1}
\left|\frac{G^{\textrm{(eff)}}Mr/6F_{\textrm{R}}^{\textrm{(b)}}}{4G^{\textrm{(eff)}}M/
3F^{\textrm{(b)}}r}\right|=\left|\frac{F^{\textrm{(b)}}}{8F_{\textrm{R}}^{\textrm{(b)}}}
\right|r^{2}\ll1
\end{align}
and
\begin{align}\label{aprox2}
\left|\frac{F_{\textrm{T}}^{\textrm{(b)}}\mathcal{M}^{\textrm{(sur)}}(\Re)r/4F_{\textrm{R}}^{\textrm{(b)}}}
{4G^{\textrm{(eff)}}M/3F^{\textrm{(b)}}r}\right|=\left|\mathcal{B}
\frac{F^{\textrm{(b)}}}{F_{\textrm{R}}^{\textrm{(b)}}}\right|r^{2},
\end{align}
where we have defined
\begin{align}\label{beta}
\mathcal{B}\equiv \frac{\mathcal{M}^{\textrm{(sur)}}(\Re)}{M}\frac{3F_{\textrm{T}}^{\textrm{(b)}}}{16G^{\textrm{(eff)}}}.
\end{align}
In definition (\ref{beta}), the value of the first fraction is of
the order of the magnitude of a surface mass per the total mass of
the body which is usually less than $1$. Thus, if
$3F_{\textrm{T}}^{\textrm{(b)}}/16G^{\textrm{(eff)}}$ does~not get
infinite or large values, one can neglect the fourth term in
solution (\ref{psi-1})\footnote{In fact, one must ensure that this
term is negligible for models under consideration by exact
inspection, however, we omit this term for simplification
purposes.}. Note that the fourth term in solution (\ref{psi-1})
already vanishes for minimal models, for which $F_{\textrm{T}}=0$.
Finally, the form of the potential $\Psi$ reads
\begin{align}\label{psi-final}
\Psi(r)\simeq-\frac{4G^{\textrm{(eff)}}M}{3F^{\textrm{(b)}}r}+
\frac{3\mathcal{F}^{\textrm{(b)}}M}{8\pi F^{\textrm{(b)}}r}=-
\left[\frac{4}{3}G+\frac{1}{8\pi}\left(\mathcal{F}^{\textrm{(b)}}
+4\mathcal{F}_{\textrm{T}}^{\textrm{(b)}}T^{\textrm{(b)}}\right)\right]
\frac{M}{F^{\textrm{(b)}}r}.
\end{align}

Comparing $\Psi(r)$ with the Newtonian potential,
$-G^{(\textrm{N})}M/r$, gives
\begin{align}\label{G}
G=\frac{3}{4}F^{\textrm{(b)}} G^{\textrm{(N)}}-\frac{1}{8\pi}\left(\frac{3}{4}\mathcal{F}^{\textrm{(b)}}
+\mathcal{F}_{\textrm{T}}^{\textrm{(b)}}T^{\textrm{(b)}}\right),
\end{align}
which shows that the deviation from the corresponding result of
$f(R)$ gravity~\cite{solar1} is the appearance of the second and
third terms that depend on the background matter density.
Nevertheless, the appearance of the trace--dependent terms is~not
a new result. Actually, it is reminiscent of the Palatini
formulation of $f(R)$ gravity theories, wherein the trace of the
corresponding field equations reads~\cite{olmo}
\begin{align*}
RF(R)-2f(R)=8\pi G T,
\end{align*}
where $R$ is the metric--independent Ricci scalar. This equation
for a well--defined function of $f(R)$ provides an algebraic
equation for the Ricci scalar, and therefore, one can conclude
that $R=R(T)$ and $F=F(T)$. This dependence has some interesting
results in the solar system applications. In the weak--field limit
of the Palatini formulation of these theories, for a pressureless
fluid, one obtains~\cite{olmo}
\begin{align*}
g_{tt}=-\frac{1}{\varphi(T)}\left( 1-\frac{2GM(r)}{r}\right)e^{2(\psi(r)-\psi_{0})},
\end{align*}
where $M(r)$ and $\psi(r)$ are some functions of the radius $r$,
$\psi_{0}$ is a constant and $\varphi(T)\equiv F(0)/F(T)$. As is
obvious, $g_{tt}$ has been modified by a trace--dependent
coefficient as our result(\ref{psi-final}).

In our case, the potential $\Phi(r)$ is also attained from
equation (\ref{perturbed-field2}), by using solutions
(\ref{solution-R1}) and (\ref{psi-final}) outside the spherical
body, i.e.,
\begin{align}\label{phi-final}
\Phi(r)\simeq\frac{2G^{\textrm{(eff)}}M}{3F^{\textrm{(b)}}r}-
\frac{3\mathcal{F}^{\textrm{(b)}}M}{8\pi F^{\textrm{(b)}}r}=
\left[\frac{2}{3}G+\frac{1}{8\pi}\left(-\mathcal{F}^{\textrm{(b)}}
+\frac{2}{3}\mathcal{F}_{\textrm{T}}^{\textrm{(b)}}T^{\textrm{(b)}}\right)\right]
\frac{M}{F^{\textrm{(b)}}r}.
\end{align}
One can easily check that solutions (\ref{psi-final}) and
(\ref{phi-final}) satisfy equation (\ref{perturbed-field3})
outside the body. Hence, in $f(R,T)$ gravity, the PPN gamma
parameter, which is related to the potential $\Phi$ via
$\Phi=\gamma G^{\textrm{(N)}}M/r$~\cite{will}, is obtained to be
\begin{align}\label{gamma}
\gamma^{(\textrm{f(R,T)})}=\frac{1}{2}-\frac{3\mathcal{F}^{\textrm{(b)}}}{16\pi G^{\textrm{(N)}} F^{\textrm{(b)}}}.
\end{align}
That is, in $f(R,T)$ gravity, we generally have a running PPN
parameter that depends on both the background matter density and
the Ricci scalar . Thus, $f(R,T)$ gravity may has some chances to
be made consistent with the solar system experiments by
constructing some plausible models.

In equation (\ref{gamma}), by setting
$\mathcal{F}^{\textrm{(b)}}=0$, one recovers the corresponding
result obtained for $f(R)$ gravity as in Ref.~\cite{solar1}.
However, in the literature~\cite{olmo1,olmo2}, it has been
indicated that applying a scalar--tensor representation for $f(R)$
gravity leads to a significant effect in its corresponding PPN
parameter. Actually, the corresponding result
$\gamma^{(\textrm{f(R)})}=1/2$ of Ref.~\cite{solar1} is valid when
the condition $|m_{\textrm{f(R)}}^{2}|r^{2}\ll1$ is held. This
condition, in a scalar--tensor representation of $f(R)$ gravity,
denotes a light scalar field (i.e., $|m^{2}_{\textrm{f(R)}}|\ll1$)
with long interaction ranges. On the other hand, one can obtain
$\gamma^{(\textrm{f(R)})}\simeq1$ by setting the condition
$|m_{\textrm{f(R)}}^{2}|r^{2}\gg1$, which is translated to as the
appearance of a heavy scalar field (i.e.,
$|m^{2}_{\textrm{f(R)}}|\gg1$) with short interaction ranges.
Therefore, depending on the mass of the scalar degree of freedom
related to the curvature (which is model--dependent), one can get
the desired result for this parameter, that is, the PPN parameter
in $f(R)$ gravity is~not also constrained to the value $1/2$.

In the following, we consider the $\gamma^{(\textrm{f(R,T)})}$ parameter for general
minimal\footnote{We temporarily relax the constraint relation (\ref{source}).} power law $f(R,T)$ models.\\

$\mathbf{General~Minimal~Power~Law~Case}~f(R,T)=c_{\textrm{R}}R^{n}+c_{\textrm{T}}T^{m}$\\

These models have
\begin{align}\label{model2-1}
&F=nc_{\textrm{R}}R^{n-1},~~~~~~~~F_{\textrm{R}}=n(n-1)c_{\textrm{R}}R^{n-2},\nonumber\\
&\mathcal{F}=mc_{\textrm{T}}T^{m-1},~~~~~~\mathcal{F}_{\textrm{T}}=m(m-1)c_{\textrm{T}}T^{m-2},\nonumber\\
&F_{\textrm{T}}=0,~~~~~~~~~~~~~~~~\mathcal{F}_{\textrm{R}}=0,
\end{align}
where $c_{\textrm{R}}$ and $c_{\textrm{T}}$ are two coupling
constants. The potential $\Psi$, for these models, is
\begin{align}\label{model2-2}
\Psi(r)=\frac{-4GM}{3nc_{\textrm{R}}R^{\textrm{(b)}(n-1)}r}+c_{\textrm{T}}\frac{m(1-4m)T^{\textrm{(b)}
(m-1)}}{24\pi n c_{\textrm{R}}R^{\textrm{(b)}(n-1)}},
\end{align}
which, by comparing it with the Newtonian potential, we reach at
the following result
\begin{align}\label{model2-3}
G=
\left\{
\begin{array}{l}
n^{-1}c_{\textrm{R}}R^{\textrm{(b)}(1-n)}\left(\frac{3}{4}G^{\textrm{(N)}}+
\frac{c_{\textrm{T}}}{8\pi}m(\frac{1}{4}-m)\rho^\textrm{(b)(m-1)}\right),~~~~~~\mbox{for odd $m$}\\\\
n^{-1}c_{\textrm{R}}R^{\textrm{(b)}(1-n)}\left(\frac{3}{4}G^{\textrm{(N)}}-
\frac{c_{\textrm{T}}}{8\pi}m(\frac{1}{4}-m)\rho^\textrm{(b)(m-1)}\right),~~~~~~\mbox{for even $m\neq0$}.
\end{array}
\right.
\end{align}
For positive values of $c_{\textrm{T}}$ and $m$, the former cases
can lead to $G<G^\textrm{(N)}$, while the latter ones can lead to
$G>G^\textrm{(N)}$. The PPN gamma parameter for these models
becomes
\begin{align}\label{model2-4}
\gamma=
\left\{
\begin{array}{l}
\frac{1}{2}-\left(3c_{\textrm{T}}m\rho^{\textrm{(b)(m-1)}}\right)/\left(16\pi n G^{\textrm{(N)}}
c_{\textrm{R}}R^{\textrm{(b)}(n-1)}\right),~~~~~~\mbox{for odd $m$}\\\\
\frac{1}{2}+\left(3c_{\textrm{T}}m\rho^{\textrm{(b)(m-1)}}\right)/\left(16\pi n G^{\textrm{(N)}}
c_{\textrm{R}}R^{\textrm{(b)}(n-1)}\right),~~~~~~\mbox{for even $m\neq0$},
\end{array}
\right.
\end{align}
where $n\neq0$. Therefore, for positive values of
$c_{\textrm{T}}$, $m$, and $n$, only for even $m$, there is a
possibility to obtain $\gamma^{\textrm{(f(R,T))}}\gtrsim1$. Note
that, for consistency with the solar system experiments, one must
get the PPN parameter of the observations, i.e.,
$\gamma^{\textrm{(obs)}}=1+(2.1\pm2.3)\times
10^{-5}$~\cite{ppn1,ppn2}.

As a special case, the model $f(R,T)=R+c_{1}\sqrt{-T}$ has
$F^{\textrm{(b)}}=1$, $F_{\textrm{R}}^{\textrm{(b)}}=0$,
$\mathcal{F}^{\textrm{(b)}}=-c_{1}/(2\sqrt{\rho^{\textrm{(b)}}})$,
and
$T^{\textrm{(b)}}\mathcal{F}_{\textrm{T}}^{\textrm{(b)}}=-\mathcal{F}^
{\textrm{(b)}}/2$. Therefore, in this case, the field equations
become
\begin{align}
&\nabla^{2}\Psi+\frac{1}{2}R^{\textrm{(s)}}=
\left(8\pi G+\mathcal{F}^{\textrm{(b)}}\right)\rho^{\textrm{(s)}},\label{perfield-1}\\
&-\Psi''+\frac{2}{r}\Phi'-\frac{1}{2}R^{\textrm{(s)}}
=-\frac{1}{2}\mathcal{F}^{\textrm{(b)}}\rho^{\textrm{(s)}}
,\label{pertfield-2}\\
&\frac{1}{r}\Phi'-\frac{1}{r}\Psi'+\frac{2}{r^{2}}\Phi
-\frac{1}{2}R^{\textrm{(s)}}=-\frac{1}{2}\mathcal{F}^{\textrm{(b)}}
\rho^{\textrm{(s)}},\label{perfield-3}
\end{align}
with the solutions
\begin{align}
\Psi=-\Phi=-\frac{(32\pi G-c_{1}\rho^{\textrm{(b)}-1/2})M}{32\pi r},
\end{align}
which holds outside the spherical body. Here, one can recover the
GR solution by setting $c_{1}=0$. Also, we have
\begin{align}
G=G^{(N)}+\frac{c_{1}}{32\pi \sqrt{\rho^{\textrm{(b)}}}},
\end{align}
and as a result $\Psi=-\Phi=-G^{\textrm{(N)}}M/r$, which yields
$\gamma=1$.

\section{conclusions}\label{conclusions}
We have extended the cosmological solutions of $f(R,T)$ gravities
in a homogeneous and isotropic FLRW space--time. We consider the
minimal theories of type $g(R)+h(T)$ that primarily were studied
in our work~\cite{phsp}, however, the results indicate that they
deserve more consideration. Our studies are based on the
phase--space analysis (the dynamical system approach) via defining
some dimensionless variables and parameters. By respecting the
conservation of the energy--momentum tensor, we have shown that
the functionality of $h(T)$, in the minimal models, must be
$h(T)=c_{1}\sqrt{-T}$. In the previous work, we obtained the
results for the density parameters, the effective equation of
state, and the scale factor. We found that there can be a
consistent cosmological sequence including radiation, matter, and
acceleration expansion eras.

In this work, to check some other cosmological aspects of these
theories, we have considered a few more cosmological parameters
written in terms of the dimensionless variables that are suitable
in the dynamical system procedure. These parameters/quantities are
the equation of state of the dark energy, the Hubble parameter and
its inverse, the coincidence parameter and its variation with
respect to the time, the weight function, the deceleration and the
jerk and the snap parameters. We have presented the corresponding
equations of these quantities in terms of the defined
dimensionless variables and then have considered them numerically.
In particular, we have investigated two general theories of type
$R+\alpha R^{-n}+\sqrt{-T}$, especially for $n=-0.9$ and
$n=-1.01$, and $R[\log{(\alpha R)}]^{q}+\sqrt{-T}$, especially for
$q=1$ and $q=-1$. The Hubble parameter and its inverse have shown
similar features for these four models. Based on the chosen
initial values for each theory, from their numerical diagrams, we
have found that the former theory with $n=-0.9$ respects the
available observational data. The diagram of $H^{-1}$ for this
theory indicates that the ratio of the size of the universe in the
matter--dominated era with respect to the present era is about
$10^{-6}$, which is an acceptable value. This ratio for the rest
of these models is about $10^{-3}$ which is far from its expected
value. The numerical plots for the coincidence parameter reveal
that the matter and the dark energy densities have become of the
same order twice, in the early and late times, and hence this
coincidence is~not a unique event. Also, they show that there is a
peak in the plots, which means that the dark energy density has
never been zero. The diagrams of $dr^{\textrm{(mD)}}/dN$
illustrate that the coincidence parameter increases up to zero in
the late times, and its rate is about $-0.6$ at the present.
Furthermore, the diagrams of the coincidence parameter and its
rate function indicate that, in the early times, the matter
density had been smaller than the dark energy density, and then it
grew over the dark energy density, and today the dark energy
density is larger than it again.

All the density parameters (except the spatial curvature density)
are affected by the weight function, and hence it is wise to know
its behavior. In this respect, it must satisfy the following
conditions:
\begin{itemize}
\item[(i)]
If this function gets negative values, the corresponding density
parameters will become negative, that is, of less physical
interest.
\item[(ii)]
In the deep matter era, GR should be a limiting solution, i.e.,
when $g(R)\sim R$, then $F(R)\sim1$ in those times.
\end{itemize}
We have found that, for the theory $R+\alpha R^{0.9}+\sqrt{-T}$,
these two conditions are hold, however, for the other theories, it
has negative values or has values far from $1$. We have also drawn
the related diagrams of the deceleration, jerk, and snap
parameters for this model (the other models have diagrams of the
same features). We have obtained that they show a good
cosmological sequence during their evolutions with the present
values $q_{0}\simeq -0.04$, $j_{0}\simeq 0.22$, and $s_{0}\simeq
-0.76$.

For all the considered cosmological quantities, we have chosen the
maximum--consistent value of the model parameter, namely,
$\alpha$, that is, the theory $R\log{(\alpha R)}+\sqrt{-T}$ with
$\alpha =1.95\times10^{92}$, the theory $R/\log{(\alpha
R)}+\sqrt{-T}$ with $\alpha=5.7\times10^{-61}$, the theory
$R+\alpha R^{1.01}+\sqrt{-T}$ with $\alpha=1.371$, and the theory
$R+\alpha R^{0.9}+\sqrt{-T}$ with $\alpha=6.46\times10^{-5}$. It
is obvious that, for the first two cases, some kind of
fine--tuning problem appears. Note that all the mentioned results
have been achieved by applying some specific initial values, and
thus further inspection via the initial values may improve the
results.

In the second part of this work, we have considered the more
plausible and simple model $R+c_{1}\sqrt{-T}$ in a non--flat
geometrical background to aim at the point of whether or not
$f(R,T)$ gravity can specify a particular sign for the spatial
curvature parameter. This model has the specific parameter
$m(r)=0$, which indicates that the formalism presented in the
first part of the work cannot be applied here. Hence, the
dynamical system equations have been achieved independently for
this model. In this respect, the dynamical system approach shows
that there is no fixed point solution denoting a closed universe.
We have plotted the diagrams of the radiation, matter, dark
energy, and spatial curvature density parameters. The diagrams
illustrate that the spatial curvature density parameter has
approximately a vanishing value, and then it increases up to the
present value $\Omega^{(\textrm{k})} \varpropto 10^{-3}$. In spite
of the lack of a fixed point denoting a closed universe, we have
set some initial values to get this kind of solution, for which we
have found that it gives the value $\Omega^{(\textrm{k})}
\varpropto -10^{-4}$ at the present.

Furthermore, this simple theory has four fixed points, three of
which are the saddle points in a three--dimensional coordinate
system. One of the fixed point denotes the radiation--dominated
era, another one specifies the matter--dominated era, the third
one indicates an era with the maximum spatial curvature density,
and, finally, the last one is a stable fixed point that determines
the dark--energy--dominated era with $w^{(\textrm{DE})}=-1/2$. For
an arbitrary value of $c_{1}$, the density parameter of the dark
energy admits $\Omega^{(\textrm{DE})}=1/c_{1}$ as a critical
value, which shows the best value for $c_{1}$ is approximately
$1$. Therefore, the terms $R$ and $\sqrt{-T}$ appear of the same
order of magnitude in the Lagrangian (in our assumed unit).
Schematically, we have demonstrated that greater values for
$c_{1}$ lead to a relatively non--dominant late--time dark energy
era, and smaller values lead to negative matter density in the
late--times. Also, we have depicted the diagrams for the dark
energy equation of state for a closed and an open universe.
Irrespective of their late time values, these two diagrams have
different features. The curve of dark energy equation of state has
either an increasing behavior toward the value $w^{(\textrm{DE})}
=-1/2$ for an open universe, or has a decreasing behavior toward
the value $w^{(\textrm{DE})}=-1/2$ for a closed universe. This
oppositional behavior can, in principle, be a distinguishable
criterion. If the evolution of $w^{\textrm{(eff)}}$, in different
cosmological eras, can be investigated from the observational
data, then this increasing (decreasing) feature will rule out the
inconsistent cosmological model.

We conclude that the theory $R+\alpha R^{0.9}+\sqrt{-T}$
relatively passes more criteria than the other considered
theories, and it deserves more accurate investigations. The
investigations have been dependent on the initial values, and
hence, more accurate investigations demand precise initial values.
However, in this work, we have investigated those $f(R,T)$ gravity
theories that are formed via the corresponding $f(R)$ gravity ones
just by adding a simple $\sqrt{-T}$ term, nevertheless, this extra
term leads to some interesting features. And yet, more interesting
models in $f(R,T)$ gravity may be found within the non--minimal
cases.

In the last part of this work, we have presented the weak--field
limit of $f(R,T)$ gravity outside a spherical body immersed into
the background cosmological fluid for an isotropic and homogeneous
background space--time. Here, we have considered a pressureless
matter. In this case, the Taylor expansion of all the functions
are performed about the current background value of the Ricci
scalar and the cosmological mass density. We have found that, in
spite of the results of $f(R)$ gravity, the field equations depend
on the value of the mass density of the cosmological matter.
Actually, in this analysis, the background cosmological fluid
plays an important role. The derivations show that the mass
parameter explicitly depends on the cosmological fluid density,
and therefore it can achieve small or large values depending on
the considered model. As a result, we have obtained the PPN gamma
parameter for $f(R,T)$ gravity, and thereby we have shown that
this parameter depends on the cosmological matter density too.
Then, we conclude that one has some chances to construct $f(R,T)$
gravity models consistent with the solar system experiments. As a
special case, we have gained the PPN parameter for general minimal
power law models and have shown that these models can accept
admissible value of the PPN parameter. And, finally, we have
obtained that the PPN gamma parameter for the model
$R+c_{1}\sqrt{-T}$ is exactly $1$.
\section*{Acknowledgments}
We thank the Research Office of Shahid Beheshti University G.C.
for the financial support.

\end{document}